\documentclass[12pt,a4paper]{article}

\usepackage{color}
\usepackage{url,graphicx,xspace}

\usepackage{amssymb}
\usepackage{hyperref}

\usepackage{ifthen}
\newboolean{commentson}         
\setboolean{commentson}{true}
 \setboolean{commentson}{false}  

\newboolean{commentsaon}         
\setboolean{commentsaon}{true}
 \setboolean{commentsaon}{false}  

\newtheorem{theoremit}{Theorem}[section]

\newenvironment{theorem}[1][]{\ifthenelse{\equal{#1}{}}
                                {\begin{theoremit}\rm}
                                {\begin{theoremit}[#1]\rm}
                              }{\end{theoremit}}

\newtheorem{exampleit}[theoremit]{Example}
\newenvironment{example}[1][]{\ifthenelse{\equal{#1}{}}
                                {\begin{exampleit}\rm}
                                {\begin{exampleit}[#1]\rm}
                              }{\end{exampleit}}

\newtheorem{definitionit}[theoremit]{Definition}

\newenvironment{df}[1][]{\ifthenelse{\equal{#1}{}}
                                  {\begin{definitionit}\rm}
                                  {\begin{definitionit}[#1]\rm}
                            }{\end{definitionit}}

\newtheorem{lemmait}[theoremit]{Lemma}

\newenvironment{lemma}[1][]{\ifthenelse{\equal{#1}{}}
                                  {\begin{lemmait}\rm}
                                  {\begin{lemmait}[#1]\rm}
                            }{\end{lemmait}}

\usepackage{pstricks}

\newcommand*{\seq}[2][n]  {{#2_{1}, \allowbreak \ldots, \allowbreak #2_{#1}}}

\newcommand*{\notmodels}{\mathrel{\,\not\!\models}}

\newcommand*{\ol}[1]{\ensuremath{\overline{#1}}}
\newcommand*{\oll}{\ensuremath{\overline{L}}}
\newcommand*{\olll}{\ensuremath{\overline{L'}}}
\newcommand*{\olk}{\ensuremath{\overline{K}}}

\newcommand{\eqlooser}{ \mathbin= }
\newcommand{\eq}{ {=} }

\newcommand*{\restrict}[2]{#1{\mid}_{ #2}}

\newcommand*{\ttt}{\ensuremath{\mathbf t}\xspace}
\newcommand*{\uuu}{\ensuremath{\mathbf u}\xspace}
\newcommand*{\fff}{\ensuremath{\mathbf f}\xspace}

\newcommand{\Tr}{\ensuremath{\mathord{T\!r}}\xspace}

\renewcommand*{\P}{\ensuremath{{\cal P}}\xspace}
\newcommand*{\R}{\ensuremath{{\cal R}}\xspace}
\newcommand*{\T}{\ensuremath{{\cal T}}\xspace}
\newcommand*{\F}{\ensuremath{{\cal F}}\xspace}
\newcommand*{\I}{\ensuremath{{\cal I}}\xspace}
\newcommand*{\V}{\ensuremath{{\cal V}}\xspace}
\newcommand*{\hp}{\ensuremath{(P,{\cal T})}\xspace}

\usepackage{latexsym} 

\usepackage{bm} 

\newcommand*{\psip}{\ensuremath{\Psi_{P}}\xspace}
\newcommand*{\psipower}[1]{\ensuremath{\Psi_{P}^{#1}(\emptyset)}\xspace}
\newcommand*{\PtI}{\ensuremath{{P/\!_t I}}\xspace}
\newcommand*{\PtuI}{\ensuremath{{P/\!_{tu} I}}\xspace}

\newcommand*{\psipm}{\ensuremath{\Psi_{P/M_0}}\xspace}
\newcommand*{\psipmpower}[1]{\ensuremath{\Psi_{P/M_0}^{#1}(\emptyset)}\xspace}
\newcommand*{\PMtI}{\ensuremath{{(P/M_0)/\!_t I}}\xspace}
\newcommand*{\PMtuI}{\ensuremath{{(P/M_0)/\!_{tu} I}}\xspace}
\newcommand*{\PMt}[1]{\ensuremath{{(P/M_0)/\!_{t} #1}}\xspace}
\newcommand*{\PMtu}[1]{\ensuremath{{(P/M_0)/\!_{tu} #1}}\xspace}

\newcommand{\comment}[1]
{\ifthenelse{\boolean{commentson}}
   {{\par\noindent\mbox{}{\small\blue[ *** #1 ]\par}\noindent\par}}{}}

\newcommand{\commenta}[1]
{\ifthenelse{\boolean{commentsaon}}
   {{\par\noindent\mbox{}{\small\green[ *** #1 ]\par}\noindent\par}}{}}

\begin{document}



\title{Hybrid Rules with Well-Founded Semantics}


\author{
W\l odzimierz Drabent%
\thanks{
Institute of Computer Science,
         Polish Academy of Sciences,
         ul.\ Ordona 21,
         Pl -- 01-237 Warszawa, Poland.
}
\thanks{
Department of Computer and Information Science,
Link\"oping University, S 581\,83 Link\"oping, Sweden,
\texttt{wdr@ida.liu.se},
\texttt{jmz@ida.liu.se}.
}
\and
\addtocounter{footnote}{-1}
Jan Ma\l uszy\'nski%
\footnotemark
} 

\date{March 13, 2009}

\maketitle


\begin{center}
{\bf Submitted for publication}
\end{center}

\begin{abstract}
A general framework is proposed for
integration of rules and external first order theories.
It is based on the well-founded semantics of normal logic programs
and inspired by ideas of Constraint Logic Programming (CLP)
and constructive negation for logic programs.
Hybrid rules are normal clauses extended with constraints in the bodies;
constraints are certain formulae in the language of the external theory.
A hybrid program is a pair of a set of hybrid rules and an external
theory.  
Instances of the framework are obtained by specifying the class of
external theories, and the class of constraints. 
An example instance is integration of (non-disjunctive) Datalog with
ontologies formalized as description logics.

The paper defines a declarative semantics of hybrid programs 
and a goal-driven formal operational semantics.  The latter
can be seen as a generalization of SLS-resolution.  It provides a basis
for hybrid implementations combining Prolog with constraint solvers.
Soundness of the operational semantics is proven.  Sufficient conditions
for decidability of the declarative semantics, and for completeness of
the operational semantics are given.
\end{abstract}


\commenta{Was:
    integration of rules and ontologies,
    well-founded semantics,
    constructive negation,
    Semantic Web reasoning,
    knowledge representation.
    }

\section{Introduction}

This paper presents an approach  to
integration of   normal logic programs under the well-founded semantics
with external first-order theories. The problem is motivated by
the discussion about the rule level of the Semantic Web.%

It is often claimed %
that rule-based applications  need 
non-monotonic reasoning  for  handling  negative information.
As the issue of non-monotonic reasoning and negation 
 was thoroughly investigated in the context of logic programming
(see e.g.\ \cite{AptB94} for a survey of classical work
on this topic), it would be desirable to build-up on this 
expertise.  Well-established
 formal  semantics:   the answer set semantics \cite{BG94},
the well-founded semantics \cite{Gelder88},
and the 3-valued completion semantics of Kunen \cite{Kunen87}
 provide theoretical 
foundations for existing logic programming systems. 
On the other hand, applications refer usually
to  domain-specific knowledge, that is  often supported
by specific  reasoning or computational mechanisms.

Domain-specific variants of logic programs are handled 
within the constraint logic programming framework CLP \cite{CLPhandbook2006}.
In CLP  the concept of constraint domain makes
it possible to extend the semantics of pure logic programs
and to use domain-specific constraint solvers for sound
reasoning. Classical CLP does not support non-monotonic reasoning,
but  integration of both paradigms is discussed by 
some researchers
 (see e.g.\ \cite{Stuckey95,DixS98,Fages97}).
Special  kind of domain-specific knowledge are 
 domain-specific terminologies, specified in a formal ontology
description language, such as OWL \cite{owl-sas@w3c}. This 
raises the issue of integration of rules and ontologies, 
which has achieved a considerable
attention (see e.g.\
\cite{Rosati05,MotikSS05,EiterIST06,Rewerse05,BruijnET08,MotikR07}
and references therein).
It is commonly assumed that an ontology
is specified as a set of axioms in (a subset of) First Order Logic (FOL),
usually in a Description Logic. A rule language, typically a variant
of Datalog, is then extended by allowing a restricted use 
of ontology predicates. The extensions considered in the literature
are mostly based on disjunctive Datalog with negation under the answer set semantics.

In contrast to that, our focus is on normal logic programs under the
well-founded semantics.
Our objective 
is to extend them in such  a way that  domain-specific
knowledge represented by a first order theory can be accessed from the
rules. The theory will be called the external theory.
\comment{
}
Going beyond Datalog makes it possible to use data structures like
lists for programming in the extended rule language.
We want to define the semantics of the extended language so that
the  existing reasoners for normal logic programs and for
the external theory can be re-used for querying the extended 
programs.
Thus, our objective is to provide a framework
for hybrid integration of normal logic programs and external theories.
Integration of Datalog rules with ontologies specified in a Description Logic
could then be  handled within the framework  as  a special case.

The choice of the well-founded semantics as the basis for our
approach is motivated by
existence of top-down query answering algorithms, which
facilitate building a query answering method for our framework. 
Notice also that the well-founded semantics and the answer set semantics
are equivalent for a wide class of programs, including stratified normal
programs.

We introduce a notion of hybrid program; such a program is 
a %
pair \hp where \T is a set of axioms
in a first order language $\cal L$ and $P$ is a set
of {\em hybrid rules}. \T and $P$  share function symbols but 
have disjoint alphabets of predicate symbols. This
reflects the intuition that domain-specific knowledge is 
shared by many applications and is  not  redefined by the applications.
A hybrid rule is a normal clause
whose body may include a formula in $\cal L$, called 
{\em the constraint} of the rule. 
We define declarative semantics
of hybrid programs as a natural extension of 
the well-founded semantics of logic programs.
It is 3-valued; the semantics of 
a program \hp is a set of ground literals over the alphabet of $P$.
The semantics is undecidable (as is the well-founded semantics for normal
programs).  However it is decidable for Datalog hybrid programs,
provided the constraints are decidable.

The operational semantics presented in this paper
is goal driven, allows non ground goals, and its internal data are possibly
non ground goals and constraints.
It combines a variant
of SLS-resolution (see e.g.\ \cite{AptB94})
with handling constraints.  The latter includes checking 
satisfiability of the constraints w.r.t.\ \T, which is assumed to be
done by a reasoner of \T.
Thus the operational semantics provides a basis for development
of implementations integrating LP (logic programming)
 reasoners supporting/approximating
 the well-founded semantics  (such as XSB Prolog \cite{XSB}) 
with constraint solvers. 
The operational semantics is sound w.r.t.\ the
declarative one, under rather weak sufficient conditions.
It is complete for Datalog hybrid programs under a certain syntactic
condition of safeness.
{\sloppy\par}

\comment{
}
\comment{
}

\comment{
}

In the special case of hybrid rules without non-monotonic negation
the rules can be seen as the usual implications of the FOL, thus as
additional axioms extending \T. 
In this case every ground atom which is true in the semantics of the
hybrid program is 
a (2-valued) logical consequence of $P\cup \T$.

\commenta{
}

\commenta{
}

\comment{
}

\comment{}

The paper is organized as follows. Section~\ref{prel} gives an
(informal) introduction  to the well-founded semantics of normal logic
programs, 
and presents the notion of constraint used in this paper. Basic
ideas of Description Logics  and their use for defining ontologies
and ontological constraints are briefly discussed.
Section~\ref{sec:main} gives a formal presentation of the syntax 
and declarative semantics
 of the generic language of hybrid rules, parameterized
by the constraint domain. 
Section~\ref{sec:main.operational} introduces the operational semantics;
then soundness and completeness results
relating the declarative semantics and the operational semantics
are stated and proven.
The last two sections contain discussion of  related work
and  conclusions.
A preliminary, abbreviated version of this work appeared as
\cite{DrabentM07}.

\section{Preliminaries}
\label{prel}
\subsection{Normal logic programs and the well-founded semantics}
\label{sec:wfs}

In this work we use the standard terminology and notation of logic
programming (see e.g.\ \cite{Apt-Prolog}).

The language of hybrid rules will be defined as an extension
of {normal logic programs}. We assume
that the programs are built over a first-order alphabet 
including   a set $\P_R$ of predicates,
a set $\V$ of variables and a set $\F$ of function symbols
with different arities  including  a non-empty
set of symbols of arity 0, called {\em constants}.

Atomic formulae (or {\em atoms}) and {\em terms}
 are built in a usual way. A {\em literal}
is  an atomic formula ({\em positive literal}\/) or a negated atomic
formula ({\em negative literal}\/). A literal  (a term) not including
variables is called {\em ground}. 

A {\bf normal logic program}  $P$ is a finite set of rules of the form
$$ H \leftarrow B_1,...,B_n  \qquad  \mbox{ where } n\geq 0$$
where $H$ is an atomic formula, and $B_1,...,B_n$ are {literals}.
The rules are also called {\em normal clauses}.
The rules with empty bodies ($n=0$) are called {\em facts}
 or {\em unary clauses}; they are usually written without $\gets$.
A normal clause is called {\em definite clause} iff all literals
of its body are positive. A {\em definite program} is a finite set 
of definite clauses.
In this paper, a {\bf Datalog} program is a normal logic program
with \F being a finite set of constants.

The {\bf Herbrand base} ${\cal H}_P$ is the set of all ground atoms
built with the predicates, constants, and function symbols of $P$.
For a subset $S\subseteq {\cal H}_P$, by $\neg S$ we denote
the set of negations of the elements of $S$, \,
$\neg S = \{\, \neg a\mid a\in S\,\}$.
A ground instance of a rule $R$ is a rule $R'$ obtained by replacing
each variable of $R$ by a ground term over the alphabet.
The set of all ground instances of the rules of a program $P$
will be denoted $ground(P)$. 
Notice that in the case of Datalog $ground(P)$ is a finite set
of ground rules.

A {\em 3-valued Herbrand interpretation} (shortly -- {\bf interpretation})
$\cal I$ of $P$ is a subset of
${\cal H}_P \cup\neg{\cal H}_P$ such that for no  ground atom $A$ 
both $A$ and $\neg A$ are in $\cal I$. 
Intuitively, the set $\cal I$ assigns the truth value \ttt (true)
to all its members. Thus $A$ is false (has the truth value \fff)
in $\cal I$ iff $\neg A \in \cal I$,
and $\neg A$ is false in $\cal I$ iff $A\in\cal I$.
If $A\not\in\cal I$ and $\neg A\not\in\cal I$ then the truth value 
of $A$ (and that of $\neg A$) is \uuu (undefined).
This is in a natural way generalized to non ground atoms and non
atomic formulae (see e.g.\ \cite{AptB94}).
 An interpretation
 $\cal I$ is a model of a formula $F$
(which is denoted by $\I\models_3 F$)
iff  $F$ is true in  \I.

As usual, a {\em 2-valued Herbrand interpretation} is a subset of ${\cal H}_P$.
It assigns the value \ttt to all its elements and the value \fff to all
remaining elements of the Herbrand base.
It is well known that any definite program $P$ has
a unique least\footnote{in the sense of set inclusion.}
2-valued Herbrand model. We will denote it ${\cal M}_P$. 
A normal program may not have the least Herbrand model.

The well-founded semantics of logic programs
\cite{Gelder88} assigns to every program $P$
a unique (three valued) Herbrand model, called the {\em well-founded}
model of $P$. Intuitively, the facts of a program should be true,
and the ground atoms which are not instances of the head of
any rule should be false. This information can be used to reason
which  other atoms must be true and which must be false in any 
Herbrand model. Such a reasoning gives in the limit the well-founded
model, where the truth values of some atoms may still be undefined.
The well-founded semantics has several equivalent 
formulations. We briefly sketch here a definition following that of
\cite{FerrandD93}.
\comment{
}

While defining the well-founded model,
for every predicate symbol  $p$ we will treat $\neg p$ as a new
distinct predicate symbol.
A normal program can thus be treated as a
definite program over Herbrand base 
$ {\cal H} \cup \neg {\cal H}$.
A 3-valued interpretation over $ {\cal H}$ can be treated as a 2-valued
interpretation over $ {\cal H} \cup \neg {\cal H}$.

Let $I$ be such an interpretation ($I\subseteq {\cal H}\cup\neg{\cal H}$).
We define two ground, possibly infinite, definite programs $\PtI$ and
$\PtuI$. 
For a given program $P$,
$\PtI$  is the ground instantiation of  $P$  together with ground unary
clauses that show which negative literals are true in $I$.
\[
\PtI = ground(P) \cup  \{\, \neg A \mid \neg A \in I \,\} 
\]
$\PtuI$  is similar but all the negative literals that are true or
undefined  %
in $I$  are made true here:
\[
\PtuI = ground(P) \cup  \{\, \neg A \mid  A \not\in I,\:  A \in{\cal H} \,\}
\]

Now we define an operator $\psip(I)$ which produces a new 
Herbrand interpretation of $P$:
$$
\psip(I) = (
{\cal M}_\PtI  \cap {\cal H}
	    )  \cup  \neg ( {\cal H} \setminus
	{\cal M}_\PtuI
			   )
$$
It can be proved that the operator is monotonic;
$\psip(I)\subseteq\psip(J)$ whenever $I\subseteq J$.
Its least fixed point is called the {\bf well-founded
model} $WF(P)$ of program $P$.  For some countable ordinal $\alpha$
we have $WF(P) = \psipower\alpha$.

The following example shows a simple Datalog  program and its 
well-founded model.

\begin{example}
\label{ex:wf}
A two person game consists in moving a token between vertices of 
a directed graph. Each move consists in traversing one  edge from
the actual position. Each of the players in order makes one move.
The graph is described by a database of facts $m(X,Y)$ corresponding
to the edges of the graph.
A position $X$ is said to be a {\em winning  position} $X$  if
 there exists a move  from $X$ to a position $Y$ which is 
a losing (non-winning) position:
$$ w(X) \leftarrow m(X,Y), \neg w(Y)$$
Consider the graph
\vspace{-\medskipamount}
\[
\begin{array}[b]{ccccccc}
&&&&  d & \to & e \\
&&&&  \uparrow && \downarrow \\
b & \leftrightarrow & a & \to &
  c & \to & f
\end{array}
\]
and assume that it is encoded by the facts $m(b,a),m(a,b),\ldots,m(e,f)$ of the
program. 
The winning positions are $e,c$. The losing positions are 
$d, f$. Position $a$ is not a losing one since the player
has an option of moving to $b$ from which the partner can only 
return to $a$. This intuition is properly reflected by
the well-founded model of the program, 
it contains the following literals with the predicate symbol~$w$:
$w(c),w(e), \neg w(d), \neg w(f)$.

A non Datalog version of this example with an infinite graph is presented in 
\cite{drabent93-lpnmr}.
\end{example}

\subsection{External theories}
In this section we discuss logical theories to be integrated with logic
programs. 

\subsubsection{Constraints}
\label{sec:consd}
Our objective is to define a general framework for extending normal
logic programs, which, among others, can also be used for
 integration of Datalog rules with ontologies. Syntactically, the
clauses of a logic program are extended by adding
certain formulae of a certain logical theory.
The added formulae will be called 
{\bf constraints}.
We use this term due to similarities with constraint logic programming
\cite{CLPhandbook2006}.

We will consider a  2-valued FOL theory, called {\em external theory} or
{\em constraint theory}. 
A set of its formulae is chosen as the set of constraints.
   Our operational semantics imposes certain restrictions on the set of
   constraints.  They are introduced together with the operational semantics.
   The declarative semantics works for an arbitrary set of constraints.
The function symbols and the variables of the language of the external
theory are the same as those of the language of rules.
On the other hand, the predicate symbols of both languages are distinct.
We will call them {\em constraint predicates} and {\em rule predicates}.
We assume that the external theory is given by a set of axioms~\T
and the standard consequence relation $\vdash$ of the FOL, or equivalently
the logical consequence $\models$\,.
(Other consequence operations can be used instead;
 for instance deriving those formulae which are true in a canonical model of
 \T, 
 or in a given class of models.)
We will sometimes use \T as the name of the theory.

\commenta{}

Sometimes one deals with an external theory whose set $\F_\mathrm{c}$
of function symbols is a proper subset of the set \F of function
symbols of the rules.  For instance the external theory uses only
constants, and the rules employ term constructors
(i.e.~non constant function symbols).
In such case 
we simply extend the alphabet of the external theory so that its set
of function symbols is \F.
The modified external theory is a conservative extension of the
original one \cite{Shoenfield.logic}.
A formula without symbols from $\F\setminus \F_\mathrm{c}$ is a 
logical consequence of \T in one of them iff it is a logical
consequence in the other.
Thus such modification of the external theory is inessential;
this justifies our assumption of a common alphabet of function symbols.

\subsubsection{Ontologies and ontological constraints}
\label{sec:ont}

This section surveys some basic concepts of Description Logics (DLs)
\cite{DLHB03} and the use of DLs for specifying ontologies.
An {\em ontology}\/ may be defined as a ``specification of a
conceptualization'' \cite{gruber-ontology-def}. 
An ontology should thus  provide 
 a formal definition of the terminology to  be shared.

Desciption Logics are specific fragments of the FOL.
The syntax of a  DL is built over
disjoint
alphabets of {\em class names},
{\em property names}
and  {\em individual names}. 
    From the point of view of FOL
    they are, respectively, one and two argument predicate symbols, and
    constants.
Depending on the kind of DL,
different  constructors are provided to build class expressions
(or briefly  {\em classes}) and property expressions (or briefly 
{\em properties}).
Some DLs allow also to represent  concrete
datatypes, such as strings or integers. In that case 
one distinguishes between  individual-valued properties and 
data-valued properties.

By an  ontology we mean 
a finite set of   axioms  in some decidable  DL. 
 The axioms describe 
classes and properties of the ontology and assert facts 
about individuals and data. An ontology is thus a
DL knowledge base %
consisting of two parts:
a \emph{TBox} 
(terminology) including  class axioms and property axioms
and an \emph{ABox} (assertions) stating facts about individuals and data.
The axioms of DLs
can be seen as an alternative representation  of  FOL
formulae. 
Thus, the semantics of DLs
is defined by referring to the usual notions of interpretation and
model, and an ontology can be considered a FOL theory.

For most of decidable DLs
there exist well developed automatic reasoning techniques.
Given an ontology $\cal T$ in a DL one can use a respective
reasoner for checking  if a 
formula $C$ is a logical consequence of $\cal T$. 
If ${\cal T} \notmodels  C$ and ${\cal T} \notmodels  \neg C$
then $C$ is true in  some models of the ontology and false in some 
other models.

Ontologies are often specified in the standard 
Web ontology language OWL DL, based on the Description 
Logic $\mathcal{SHOIN}(D)$.  OWL Ontologies  can be
seen as set of axioms in this DL. 

OWL DL class axioms make it possible to state class equivalence
$A \equiv C$ and class inclusion $A \sqsubseteq C$, where $A$
is a class name and $C$ is a class expression. Class expressions are
built from class names using constructors, such as $\top$ (the universal
concept), $\bot$ (the bottom concept), intersection,
union and complement. Classes can also be described by 
direct enumeration of  members  and by {\em restrictions }
on properties (for more details see \cite{owl-sas@w3c}).

Property axioms make it possible to state  inclusion and
equivalence of properties, specify the domain and the range of a property,
state that a property is symmetric, transitive, functional, or inverse functional.

OWL DL assertions indicate members of classes and properties. Individuals
are referred to by individual names. %
It is possible to declare that   given individual names represent the same
individual  or that each of them represents a different individual.

The following example using  some expressive constructions of OWL DL
will be used in the sequel to discuss how integration of Datalog
with OWL DL ontologies  is achieved in our framework.
\begin{example}
\label{ex:ont}
In some research area an author of at least  3 books
is considered an expert. An OWL DL ontology referring
to this research area has   classes
$Author$ and  $Book$, and  a property ${\it isAuthorOf}$ with 
domain $Author$ and range $Book$. 
The class $Expert$
can now be defined using OWL DL cardinality 
restriction:\footnote{In the Manchester OWL Syntax \url{http://www.co-ode.org}}
$$Expert \ \equiv \  {\it isAuthorOf} \ \mbox{\bf min} \ 3\ {\it Book}$$

The property  ${\it isAuthorOf}$ has  the inverse property ${\it hasAuthor}$. 
The following class expression defines the class of authors which
co-authored a book with a given author $X$ (e.g.\ $smith$)
$${\it (isAuthorOf\ {\bf some}\ (hasAuthor \ {\bf value} \ X)) } $$

All individuals of class $Book$ which appear in the ontology are
declared as distinct.  The ontology 
states that the individuals  ${\it johns}$ and $brown$
of class $Author$ are the same.  (This may happen e.g.\ due 
to a change of the name of a person). 
There are also authors $smith$ and $burns$; 
$smith$, $burns$  and ${\it johns}$ are declared to be distinct.
In addition, the ontology asserts that
${\it johns}$ is the author of  the books $b1,b2$ and $brown$ is the author of 
the books $b2,b3$. %
Thus an OWL DL reasoner will conclude
that ${\it johns}$ ($brown$) is an expert.
\end{example}

\subsection{Datalog with Constraints: Introductory Examples}

We now illustrate the idea of adding constraints to rule bodies
on two simple examples.
The intention is to give an informal introduction to the semantics of
hybrid rules. 
The first example will be used later on to accompany the formal
presentation of the declarative and operational semantics of our framework.
The second one illustrates some aspects of expressing external theories
in OWL DL.

\begin{example}
\label{ex:hyrules}

The example  describes a variant of the game from Example~\ref{ex:wf}
where the rules are subject to additional restrictions. 
Assume that  the positions of the graph  represent 
geographical locations described by  an ontology.
The  ontology  provides, among others,  the following information
\begin{itemize}

\item subclass relations  (TBox axioms):
e.g.\  $Fi \sqsubseteq E$ (locations in Finland are locations in Europe);

\item  classification of some given locations represented by constants
 (ABox axioms). For instance, assuming that the positions of
 Example \ref{ex:wf} represent locations we may have
$Fi(b)$ ($b$ is a location in Finland),\,
$E(c)$ ($c$~is a  location in Europe).
\end{itemize}

We now  add some restrictions  as
ontological constraints%
\footnote{
Symbol $\neg$ is used to denote two kinds of negation.
Within a constraint it is the classical negation of the external theory.
When applied to a rule predicate, $\neg$
 denotes nonmonotonic negation.
Thus two distinct negation symbols are not needed.
}
added to the facts $m(e,f)$ and $m(c,f)$:
\[
\begin{array}{l}
w(X) \leftarrow m(X,Y), \neg w(Y) \\[1ex]
\begin{array}[t]{@{}l}
m(b,a)\\
m(a,b)\\
m(a,c)\\
m(c,d)\\
m(d,e)\\
\end{array}
\qquad
\begin{array}[t]{l}
m(c,f) \leftarrow \neg Fi(f)\\
m(e,f) \leftarrow E(f) \\
\end{array}
\quad
\begin{array}[t]{ccccccc}
\\
&&&&  d & \longrightarrow & e \\
&&&&  \uparrow
 && \downarrow
    \makebox[0pt][l]{\raisebox{.4ex}{$\scriptscriptstyle E(f)$}} \\
b & \leftrightarrow & a & \to &
  c & 
    \mathrel{\raisebox{-1.65ex}{$
  \stackrel{\textstyle\longrightarrow}{\scriptscriptstyle \neg Fi(f) }
 $}}
 & f
\end{array}
\end{array}
\]
Intuitively, this would mean that the move from $e$ to $f$ is
allowed only if $f$ is in  Europe and the move from $c$ to $f$
-- only if $f$ is not in Finland. These restrictions may influence
the outcome of the game: $f$ will still be a losing position
but if the axioms of the ontology do not allow to conclude
that $f$ is in Europe, we cannot conclude that $e$ is a winning position.
However, we can conclude that if $f$ is not in Europe then it cannot
be in Finland. Thus, at least one  of the conditions
$E(f), \neg Fi(f)$ holds.
 If $E(f)$ then, as in Example~\ref{ex:wf},
$e$ is a winning position, $d$ is a losing one, hence $c$ is 
a winning position. On the other hand, if $\neg Fi(f)$
then the move from $c$ to $f$ is allowed,
in which case $c$ is a winning position.
Therefore $c$ is always a winning position;
$w(c)$ is considered to be a consequence of the program.
\end{example}

\begin{example}
\label{ex:mayreview}
A committee of reviewers is to be created for evaluation of the applicants 
for an opened position. A reviewer  has to be
an expert, as defined by the ontology of Example \ref{ex:ont} and
must not have a conflict of interest (\textit{coi}) with an applicant.
Persons who are co-authors of a book have \textit{coi}.
(This implies that an author of a book has \textit{coi} with himself/herself;
this applies in particular  to each expert).
Additionally,  some conflicts of interest are declared by facts.

The following rules define 
a potential reviewer $X$ for a candidate $Y$ (relation $mayreview$).
Two constraints are used: $Expert(X)$ and
$ ({\it isAuthorOf}\ \linebreak[3]{\bf some}\linebreak[3]\ 
({\it hasAuthor} \ {\bf value} \ X))(Y)$.
They refer to the ontology of Ex.\  \ref{ex:ont}.
\[
\begin{array}{l}
mayreview(X,Y) \leftarrow Expert(X), \neg coi(X,Y) \\[.5ex]
coi(X,Y) \leftarrow ({\it isAuthorOf}\ {\bf some}\ ({\it hasAuthor} \ {\bf value} \ X))(Y)\\[.5ex]
coi({\it johns}, burns)

\end{array}
\]
The intention is to query the rules and the ontology  
for checking if a given 
person may be a reviewer for a given candidate.
Consider the individual $ {\it johns} $ of   Example \ref{ex:ont} 
and check if  she might be appointed a reviewer for 
some of the people named in the ontology.
An OWL DL reasoner can  check 
 that ${\it johns}$ is an expert and that she has the conflict of
 interest with herself, i.e.\ with ${\it johns}$ alias $brown$.
The conflict of interest with $burns$ is stated explicitly.
So ${\it johns}$ cannot be appointed a reviewer for herself 
and for $burns$.

To check if ${\it johns}$ has the conflict of interest with $smith$
one has to refer to the ontology for checking if 
they co-authored a book.
If this is confirmed by the reasoner
(e.g.\ when the ontology asserts that both 
 ${\it johns}$  and $smith$ are authors of $b1$) then
$coi({\it johns},smith)$ is true and ${\it johns}$ cannot be a reviewer for
$smith$. 
If non-existence of any co-authored book follows from the ontology,
then $coi({\it johns},smith)$ is false and ${\it johns}$ can be a reviewer.
\comment{}
Otherwise%
\footnote{
They are co-authors in some models of the ontology, and are not in some others.
}
${\it johns}$ may be a reviewer for $smith$ under the 
condition that they did not co-authored a book.
This constraint  should be returned in the answer to the query.

\comment{
}
\end{example}

An example employing non-nullary function symbols is given in 
\cite{DrabentHM07iclp}.
The  semantics of hybrid programs presented below formalizes
the intuitions presented in the examples of this section.

\section{Integration of  rules and external theories}
\label{sec:main}

This section defines the syntax and the (declarative) semantics 
 of hybrid programs,  integrating normal rules with
 first-order  theories.
The general principles discussed here apply
in a special case  to integration of Datalog with ontologies specified
in Description Logics.

\subsection{Syntax}

 We consider  a first-order alphabet
including, as usual, disjoint alphabets of  predicate symbols   $\P$, 
function symbols $\F$ (including a set of constants) and variables $\V$.
We assume that $\P$ consists of two disjoint
sets $\P_R$ ({\em rule predicates})  and $\P_C$ 
({\em constraint predicates}). 
The atoms and the literals constructed with these predicates will
respectively be called {\em rule atoms} ({\em rule literals}) and
{\em constraint atoms} ({\em constraint literals}).
    We will combine rules over alphabets $\P_R,\, \F,\, \V$
    with an external theory \T over  $\P_C,\, \F,\, \V$,
    employing constraints (a distinguished set of formulae of \T).

\begin{df}

A {\bf hybrid rule} 
(over  $\P_R,\, \P_C,\, \F,\, \V$) is
an expression of the form:
\[
H \leftarrow C,\seq L
\]
where, $n\geq0$ 
each $L_i$ is a rule literal and $C$ is a constraint
(over  $\P_C,\, \F,\, \V$); $C$ is called the {\bf constraint of the rule}.

A {\bf hybrid program}  is a pair \hp where $P$ is
a set of hybrid rules and $\cal T$ is a set of axioms
over  $\P_C,\, \F,\, \V$.
\hfill  $\Box$
\end{df}

Hybrid rules are illustrated in Example \ref{ex:hyrules}.
We adopt a convention that a constraint {\bf true}, which
is a logical constant interpreted as \ttt, is omitted. 
Usually we do not distinguish between sequences, like $\seq L$,
and conjunctions, like $L_1\land\ldots\land L_n$.
Notation $\ol L$ will be used to denote a sequence of 
rule literals
(similarly $\ol t$ a sequence of terms, etc.);
$\ol t \eq \ol u$ will denote a conjunction of equalities
$t_1 \eq u_1,\ldots,t_k\eq u_k$.

\subsection{Declarative Semantics}
\label{sec:decl.sem}

The declarative semantics of hybrid programs is defined as
a generalization of the well-founded semantics of normal programs;
it refers to the models of the external theory $\cal T$ of a
hybrid program.
Given a hybrid program \hp
we cannot  define a unique well-founded model  of $P$
since we have to take into consideration the logical values
of the constraints in the rules. However, a  unique well-founded 
model can be defined for any given model of ${\cal T}$.
Roughly speaking, the constraints in the rules are replaced by their logical
values in the model (\ttt or \fff);
then the well-founded model of the obtained logic program is
taken. 
The well-founded models are over the Herbrand universe, but
the models of \T are arbitrary.

By applying a substitution $\theta = \{x_1/t_1,\ldots,x_n/t_n\}$
to a formula $F$ we mean applying it to the free variables of $F$.
Moreover, if a bound variable $x$ of $F$ occurs in some $t_i$ ($1\leq i\leq n$)
then $x$ in $F$ is replaced by a new variable.

By a {\bf ground instance} of a hybrid rule $H \leftarrow C,\seq L$,
where $C$ is the constraint of the rule,
we mean any rule  $H\theta \leftarrow C\theta, L_1\theta,\ldots,L_n\theta$,
where 
$\theta$ is a substitution replacing the free variables of 
$H\gets C,\seq L$ 
 by ground terms (over the alphabet \F).
So the constraint $C\theta$ has no free variables, and
$H\theta, L_1\theta,\ldots,L_n\theta$ are ground literals.
By $ground(P)$ we denote the set of all ground instances of the 
hybrid rules in $P$.

\begin{df}
\label{def.declarative}
Let \hp be a hybrid program and let $M_0$ be a model of  $\cal T$. 
Let ${ P}/M_0$ be the normal  program obtained 
from $ground({P})$ by
\begin{itemize}
\item removing each rule constraint $C$ which is true in $M_0$
(i.e.\ $M_0\models C$),
\item removing each rule whose constraint $C$ is not true in $M_0$,
(i.e.\ $M_0\notmodels C$).
\end{itemize}
The well-founded model $WF({P}/M_0)$ of ${P}/M_0$ is called the {\bf well-founded model}
of $ P$ {\em based} on  $M_0$.

A formula  $F$
(over $\P_R, \F, \V\/$)
{\bf holds} (is {\em true}) in the well-founded semantics of a hybrid program
\hp (denoted $\hp\models_{\mathrm{wf}}F$)
iff  $M\models_3 F$  for each well-founded model  $M$  of \hp.
\hfill $\Box$
\end{df}

Notice that the negation in the rule literals is non-monotonic,
and the negation in the constraints is that from the external theory,
thus monotonic.

We say that $F$ is {\em false} in the well-founded semantics of \hp
if  $\hp\models_{\mathrm{wf}}\neg F$, and that $F$ is {\em undefined}
if the logical value of $F$ in each well-founded model of \hp is \uuu.
There is a fourth case:  $F$ has distinct logical values in various
well-founded models of $P$.
Formally, the semantics of \hp does not assign any truth value to such $F$.
We may say that
its truth value depends on the considered model of the external theory.
Classes  of models  in which
$F$ has a specific truth value can by  characterized by 
constraints.  Such constraints provide sufficient conditions 
for $F$ to have the specific truth value. 
They are constructed by the proposed operational semantics.

\commenta{
}

\begin{example}
\label{ex:hy-wf}
For  the hybrid program $\hp$ of Example~\ref{ex:hyrules} we have to 
consider models of the ontology~\T. For every model $M_0$  of \T
such that  $M_0 \models E(f)$ the program
 $P/M_0$ includes the fact $m(e,f)$.
The well-founded model of $P/M_0$ includes thus
the literals $\neg w(f), w(e), \neg w(d), w(c)$
(independently of whether $M_0\models Fi(f)$).

On the other hand, for every model $M_1$ of the ontology 
 such that $M_1 \models \neg Fi(f)$ the program
 $P/M_1$ includes the fact $m(c,f)$.
The well-founded model of $P/M_1$ includes thus
the literals  $\neg w(f), w(c)$
(independently of whether $M_1\models E(f)$\,).

Notice that each of the models of the ontology falls in one of
the above discussed cases.
Thus, $w(c)$ and $\neg w(f)$
hold in the well-founded semantics of the hybrid program, and 
the logical value of $w(a)$ and that of $w(b)$ is \uuu in each
well-founded
model of the program.
On the other hand  $w(e)$ and $\neg w(d)$ are  true in those well-founded
models $WF({P}/M_0)$ of $P$ for which 
the constraint $E(f)$ is true in $M_0$. 
Similarly, $\neg w(e)$  and $w(d)$  are  true in those models for which 
 $E(f)$ is false.
Thus the well-founded semantics assigns unique truth values to
$w(a), w(b), w(c)$ and  $w(f)$, but not to
$w(d)$ and $w(e)$.
The truth values of $w(d)$ and $w(e)$
can be characterized 
by additional constraints. 

\commenta{
}

\end{example}

Consider a case of hybrid rules without negative rule literals.
So the non-monotonic negation does not occur.
Such rules can be seen as implications of FOL and treated as axioms added to
\T.
For such case the well-founded semantics is
{\em compatible with FOL}
in the following sense:
For any ground rule atom $A$
if $\hp\models_{\mathrm{wf}}A$ then $P\cup \T\models A$.%
\footnote{
The reverse implication does not hold. 
As a counterexample take  $\T = \{\,\exists x.q(x)\,\}$ and
$P=\{\, p\gets q(x),r(x);\ r(x)\!\gets\,\}$.  
$P\cup \T\models p$  but  $\hp\notmodels_{\mathrm{wf}}p$,
as there exist models of \T in which each ground atom $q(t)$ is false.

We can obtain (something close to) the reverse implication 
by considering only those well-founded models which are based on
Herbrand models of \T. 
If $P\cup \T\models A$ then $M\models A$ for each well-founded model $M$ of
$P$ based on a Herbrand interpretation $M_0$ of \T.
}
We omit a detailed proof.
\commenta{
}%

\commenta{
Could a nice equivalence for safe rules be possible?
}

 As the well-founded
semantics of normal programs is undecidable, so is the well-founded
semantics of hybrid programs.
It is however decidable for Datalog hybrid programs with decidable external
theories (Section \ref{sec:decidable}).
In Section  \ref{sec:main.operational}
we show that sound reasoning is possible
(for arbitrary hybrid programs) by appropriate generalization of
SLS-resolution.
For the Datalog case 
the proposed reasoning scheme is complete 
under a certain safeness condition.

\comment{
}

\subsection{Treatment of Equality}
\label{sec:equality}

\comment{
}

In this section we discuss how equality is treated by the declarative
semantics introduced above.  The semantics is based on Herbrand models.
Thus it treats distinct ground terms as having different values. 
\begin{example}
\label{ex:noncongruent}
Consider a hybrid program \hp, where  $P = \{\,p(a)\,\}$.
Both $p(a)$ and $\neg p(b)$ hold in the well-founded semantics of \hp,
even if \T implies that $a$ and $b$ are equal.
This feature of the semantics of hybrid programs may be found undesirable.
\end{example}

We will call this phenomenon the {\em problem of two equalities}.
Below we first show that the problem is well known in constraint logic
programming (CLP) and explain how it is dealt with.
Then we discuss two more formal ways of avoiding it: 
external theories where equality satisfies Clark equality theory (CET),
and hybrid rules which are congruent w.r.t.\ a given external theory.

The problem of two equalities is familiar from CLP \cite{CLPhandbook2006},
and is not found troublesome in practice.
Most CLP implementations employ both syntactic equality
and equality of the constraint domain
\footnote{ See for instance the comment on an example constraint domain
on p.\,414 in \cite[Section 12.2]{CLPhandbook2006}.
}.
Let us denote the latter by $='$
(and use $=$ for the syntactic equality of the Herbrand domain).
Formally, let us treat $=$ as equality, and $='$ as an equivalence relation.
As an example consider CLP over arithmetic constraints
\cite{CLPhandbook2006}.
Terms $2+2$ and $4$ are distinct but denote the same number, we have
$2+2\not=4$ and $2+2='4$.
Constraint predicates treat $2+2$ and $4$ as equal.
(Formally, $='$ is a congruence of the constraint predicates:
$p(\seq t)$ iff $p(\seq u)$ whenever $t_1='u_1,\ldots,t_n='u_n$, for any
constraint predicate $p$.)
Other predicates may distinguish such terms.
This is related to using unification in the operational semantics;
unification is related to the syntactic equality.

Apparently the programmers find this feature natural and not confusing.
They are aware of dealing both with 
the Herbrand interpretation and with a non Herbrand one.
They know that the latter is employed only by constraint predicates.
They take care of distinguishing the two corresponding equalities.
For instance to express a fact that $size$ should be true for the number
$4$, a rule $size(N)\gets N\eq'4$
will be used (instead of a fact  $size(4).$).

\comment{
}

It what follows we refer to the free equality theory
(CET, Clark equality theory) \cite{Clark78}.  CET consists of equality axioms
\newlength{\mywidth}
\setlength{\mywidth}{.92\textwidth}
$$
\begin{minipage}{\mywidth}
$
\begin{array}{@{}ll}
x=x, \\
\ol x=\ol y  \to f(\ol x)=f(\ol y)\
&
\mbox{for each $f\in\F$,} \\
\ol x=\ol y  \to (p(\ol x) \to p(\ol y))
&
\mbox{for each predicate symbol $p$, including $=$,}
\end{array}
$
\end{minipage}
$$
and freeness axioms
$$
\begin{minipage}{\mywidth}
$
\begin{array}{@{}ll}
 f(\ol x)=f(\ol y) \to \ol x=\ol y  
\hspace*{.8em}
&
\mbox{for each }f\in\F, \\
f(\ol x)\neq g(\ol y)
&
\mbox{for each pair of distinct $f,g\in\F$,}
\\
x\neq t
&
\mbox{for each non variable term $t$ such that}
\\
&
\mbox{the variable $x$ occurs in $t$.}

\end{array}
$
\end{minipage}
$$
If the set \F of function symbols is finite then CET additionally
contains the weak domain closure axiom WDCA:%
\footnote
{
This axiom is needed for CET to be complete, in the sense that any closed
formula (with $=$ as its only predicate symbol)
has the same logical value in each model of CET.
Consider for instance  $\F=\{a\}$ and $\exists x (x\neq a)$.
This formula is true in some models of CET without WDCA, but false in its
(unique) Herbrand model.
}
\[
    \bigvee_{f\in\F}
    \exists \ol y\, (x = f(\ol y)).
\]
When \F contains only constants then CET reduces to 
the unique name assumption (UNA).

Assume that we have an external theory \T with equality $='$.
We say that a set of hybrid rules $P$ is {\bf congruent} for a
predicate symbol $p$ w.r.t.\ \T when
 $\T\models t_1='u_1,\ldots,t_n='u_n$ implies
\[
\hp\models_{\mathrm{wf}} p(\seq t) \mbox{ \ \ \ iff \ \ \ }
\hp\models_{\mathrm{wf}} p(\seq u),
\]
for any  ground terms $\seq t,\seq u$.
When $P$ is congruent w.r.t.\ \T for any rule predicate $p$
then we say that  $P$ is congruent w.r.t.\ \T 
(or shortly that \hp is congruent).
\comment{
}

\begin{example}
Program   $P = \{\,p(a)\,\}$ (from Ex.\ \ref{ex:noncongruent})
is not congruent w.r.t.\ any \T in which $\T\models a\neq'b$.

The hybrid program from Ex.\ \ref{ex:mayreview} 
with the fact $coi({\it johns}, burns)$ removed is congruent
\footnote{
    The unchanged program
    is not congruent,
    unless the ontology implies that ${\it johns}$ ($brown$)
    co-authored a book with $burns$.
     This is because 
     $coi({\it johns},burns)$ holds and $ coi(brown,burns)$ does not
    hold in the well-founded semantics of the program,
    but  ${\it johns} ='brown$.
},
independently from \T.
\end{example}

\begin{example}[Constructing congruent programs]
\label{ex:congruent}
Consider the program from Examples \ref{ex:hyrules}, \ref{ex:hy-wf}.
The program implies $w(c)$ and $\neg w(g)$
(formally $\hp\models_{\mathrm{wf}} w(c)$ and 
$\hp\models_{\mathrm{wf}} \neg w(g)$).
Assume that
 \T implies that $c=' g$.
For instance, the equality may be explicitly stated by an
{\tt owl:sameAs} assertion.
Informally, equality $c=' g$ is incompatible with $P$;
the rules of $P$ treat differently the objects $c,g$,
while \T states that they are equal. 
Formally,  \hp is not congruent.
{\sloppy\par}

\commenta{
}

One can modify $P$ to make it treat $c,g$ in the same way.
It is sufficient to add rules  $m(a,g)$, $m(g,d)$, and
$m(g,f)\leftarrow \neg Fi(f)$.
(We replace $c$ by $g$ in the rules of $P$).
Now  $w(c)$ and $ w(g)$ hold in the well-founded semantics of
the obtained program $(P',\T)$.
The program is congruent, provided that \T does not imply $t_1='t_2$
for any other pair $\{t_1,t_2\}\neq \{c,g\}$ of constants occurring in
the program.

We can modify $P$ to make it congruent independently from \T.
The idea is to replace (implicit) $=$ by explicit $='$. 
For instance we may replace in $P$
 the rule  $w(X) \leftarrow m(X,Y), \neg w(Y)$ by
\[
 w(X) \leftarrow X='X',\, Y='Y',\, m(X',Y'),\, \neg w(Y).
\]
The obtained program $(P'',\T)$ is congruent for $w$, w.r.t.\ any \T.
Alternatively, the rules for $m$ can be modified in a similar way to
make the program congruent for $m$.  Then the program is also
congruent for $w$ (without modifying the rule for $w$).

\comment{}

The program transformations above can be seen as usual CLP programming
tricks.

\end{example}

For congruent hybrid programs the problem of two equalities does not exist.
(Also, it does not exist for external theories without equality.)
The example above informally introduces programming techniques for
constructing congruent programs.
Now we present two simple criteria assuring that a program is congruent.
(Congruency is undecidable, like other non trivial semantic properties of
programs.)

First, if the equality $='$ of \T satisfies CET
then each program \hp is congruent.
(As then if $t,u$ are ground terms then \mbox{$t='u$} implies that the
terms are identical.) 
Apparently for this reason
 some approaches of combining rules and ontologies require that
the ontology satisfies the unique name assumption (UNA).

Another sufficient criterion is syntactic.  Program \hp is congruent if
in each rule $H\gets C,\seq L$ of $P$
all the arguments $\seq t$ of the head $H=p(\seq t)$ are variables,
and any variable occurs at most once in $H,\seq L$.
(Thus the remaining occurrences of the variable are in the constraint $C$
of the rule.)
The proof that such \hp is congruent is based on the fact that
for any model $M$ of \T if
 $\T\models t_1\eq'u_1,\ldots,t_n='u_n$ (for ground terms $\seq t,\seq u$)
then a rule $p(\seq t)\gets \ol B$ is in $P/M$ iff
 $p(\seq u)\gets \ol B$ is in $P/M$.

As an example, notice that the rule for $w$ in $P''$
(from Ex.\ \ref{ex:congruent}) satisfies the sufficient
condition, and the rules for $m$ do not.
Notice also that the condition is different from usually considered
safeness conditions (the former -- roughly speaking -- forbids certain
variable occurrences, while the latter require).

\commenta{
}

It is rather obvious how to construct programs satisfying this syntactic
restriction, provided that the set of constraints includes equalities
$t='u$ of \T.
Instead of placing a non variable term $t$ as an argument of the head of a
rule, use a new variable $x_t$ and add $x_t='t$ to the constraint of the
rule. 
Instead of writing more than one occurrences of a variable $x$ in the
rule~literals of a rule, replace each (but one) occurrence of $x$ by a new
distinct variable $x'$ and add $x'='x$ to the constraint of the rule.

\comment{
}

\section{Reasoning with hybrid rules}
\label{sec:main.operational}
Now we present a way of computing the well-founded semantics of Definition
\ref{def.declarative}.
Like in logic programming,
the task is to find instances of a given goal formula $G$ which are true
in the well-founded semantics of a given program.
Similarly to logic programming, our operational semantics is defined
in terms of search trees. 
After introducing the operational semantics we prove its soundness
and completeness, the latter for a restricted class of programs.

\subsection{Constraints for the operational semantics}
\label{sec:constr.oper}

   To construct the operational semantics we impose certain requirements on
   the external theory and the set of constraints.
   We need to deal explicitly with the
       syntactic equality $=$  and its negation.
       So we require that = is a constraint predicate symbol and the external
       theory \T includes the axioms CET (cf.\ Section \ref{sec:equality}).
   An external theory $\T'$ which does not satisfy this condition
   can be easily converted to a \T which does.
   ($\T'$ may be a theory without equality, or contain equality  $='$ not
   satisfying CET.)
   Namely \T is the union
       $\T=\T'\cup \rm CET$.
   Reasoning in such \T can be implemented employing Prolog and a reasoner
   for $\T'$ \cite{DrabentHM07}. 
   The former deals with $=$, the latter with the predicates of $\T'$ .

   The operational semantics constructs new constraints using
   conjunction, disjunction, negation, and existential quantification.
   So we require that the set of constraints is closed under these
   operations.
This imposes restrictions on the constraints.  For instance many DLs do not
allow negation of roles; for such DL a formula
of the form $r(X,Y)$ cannot be a constraint.
   The actual choice of constraints is outside of the scope of this paper.
   It depends on the 
   chosen external theory and the available reasoner for it.
   For instance, if a formula $C$ is a constraint without $=$ then the
   reasoner should be able to check whether $C$ is satisfiable in $\T'$
   (where \T and $\T'$ are as above).

\subsection{Operational Semantics}\label{sec:operational}

The operational semantics presented below is a generalization of
SLS-res\-ol\-ution \cite{Prz:JAR89},
which is extended by handling constraints 
originating from the hybrid rules.
It is based on the constructive negation approach presented 
in \cite{drabent93-lpnmr,drabent95}.
In logic programming, 
the term constructive negation stands for generalizations of negation as
failure (NAF)  (see e.g.\ \cite{AptB94}).
NAF~provides a way of checking whether a given negative goal is a
consequence of the program (under a relevant semantics).
Constructive negation, roughly speaking, finds instances of a negative
goal which are consequences.
The main contribution of the operational semantics presented here is
dealing with hybrid programs and arbitrary external theories.
The constructive negation method of \cite{drabent93-lpnmr,drabent95}
dealt with logic programs, the equality was the only constraint
predicate and CET was the constraint theory.

The operational semantics is similar to SLDNF- and SLS-resolution
\cite{lloyd87foundation,Prz:JAR89}.
For an input goal a derivation tree is constructed; its nodes are
goals.  Whenever a negative literal is selected in some node,
a subsidiary derivation tree is constructed.
So a tree of trees is obtained.

\begin{df}
\label{def.restriction...}
By the {\bf restriction} $\restrict F V$ of a formula $F$ to a set $V$ of
variables
we mean the formula $\exists x_1,\ldots,x_n F$  where  
$x_1,\ldots,x_n$ are those free variables of $F$ that are not in $V$.
By $\restrict F {F'}$ we mean $\restrict F V$, where $V$ are the free
variables of formula~$F'$.

By a 
{\bf goal} we mean a conjunction of the form  $C,\seq L$ ($n\geq0$), where
each $L_i$ is a rule literal 
and $C$ is a constraint (the {\bf constraint}  of the goal\/).
Consider a goal $G=C,\oll,p(\ol t),\olll$  and a rule
$R=p(\ol u)\gets C',\olk$, such that
no variable occurs both in $G$ and $R$.
We say that the goal 
\[
G'\  = \ \ol t\eq \ol u, C, C', \oll,\olk,\olll
\]
is {\bf derived} from $G$ by $R$, with the selected atom $p(\ol t)$, if
the constraint
$\ol t \eq\ol u, C, C'$ is satisfiable.
\end{df}

We inductively define two kinds of derivation trees: t-trees and tu-trees.
Their role is to find out when a
given goal is \ttt, or respectively when it is \ttt or \uuu.
Informally,
if a constraint $C$ is a leaf of a t-tree with the root $G$
then $C$ implies that $G$ is \ttt in the well-founded semantics of the
program. 
(More generally, the same holds if $C$ is a disjunction of such leaves.)
On the other hand, for a tu-tree we define a notion of its
{\em cross-section}.
If $\seq C$ are the constraints of the goals of a cross-section of a tu-tree
with the root $G$
then, roughly speaking, $\neg(C_1\lor\ldots\lor C_n)$ implies that 
$G$ is \fff in the well-founded semantics of the program.
A formal explanation is provided by the soundness theorem
(\ref{theo.soundness}) in the next section.

\comment{
}

For correctness of the definition (to avoid circularity)
 we assign ranks to the trees.
This is a standard technique employed in similar definitions
\cite{lloyd87foundation,Prz:JAR89,drabent95}.
In the general case ranks are countable ordinals, but
natural numbers are sufficient for a language where the
function symbols are constants.
The children of nodes with an atom selected are defined as in the standard
SLD-resolution.  The only difference is that instead of explicit unification
we employ equality constraints.  The children of nodes with a negative literal
selected are constructed employing the results of tu- (t-) trees of lower
rank.  A t-tree refers to tu-trees and vice versa.
This is basically a reformulation
of the corresponding definitions of \cite{drabent93-lpnmr,drabent95}.

\comment{
}

\comment{
}

\begin{df}[Operational semantics]\label{def.operational}
A t-tree (tu-tree)
of rank $k\geq0$ for a goal $G$ w.r.t.\ a program \hp
satisfies the following conditions.
The nodes of the tree are (labelled by) goals.  In each node
a rule literal is selected, if such a literal exists.
A node containing no rule literal is called {\bf successful},
a branch of the tree with a successful leaf is also called
{\bf successful}.
\begin{enumerate}
\item\label{positive.answer}
A constraint
 $\restrict{(C_1\lor\cdots\lor C_n)}G$ ($n\geq0$)%
\footnote{
If $n=0$ then by $C_1\lor\cdots\lor C_n$ we mean 
{\bf false}, and by $C_1\land\cdots\land C_n$ we mean
{\bf true}.
}
 is an {\bf answer}
of the t-tree if $\seq C$ are (some of the) successful leaves of the t-tree.
(It is not required that all the successful leaves are taken.)

\item\label{negative.answer}
By a {\bf cross-section} (or frontier) of a tu-tree we mean a set $F$
of tree nodes such that each successful branch of the tree has a node 
in~$F$.
Let $F$ be a cross-section of the tu-tree and ${\it CF} = \{\,C_1,\ldots\,\}$
the constraints of the nodes in $F$.  

If  ${\it CF} = \{\,\seq C\,\}$ is finite then the constraint
$\neg(\restrict{C_1}G),\ldots,\neg(\restrict{C_n}G)$ 
(the negation of $\bigvee(\restrict{C_i}G)$) is called a {\bf negative answer}
of the tu-tree.

If $\it CF$ is infinite then a constraint $C$
 which implies
$\neg(\restrict{C_i}G)$ for each  $C_i\in\it CF$
is called a {\bf negative answer}
of the tu-tree.
Moreover it is required that each free variable of $C$ is a free variable of
 $G$.
\comment{
A notation for free variables?
}

\item\label{sls.positive}
If (in the t-tree or tu-tree) the selected literal $A$ in a node $G'$ is an
atom then, 
for each rule $R$ of $P$,
a goal derived from $G'$ with $A$ selected by a variant $R'$ of $R$
is a child of  $G'$, provided such a goal exists.
Moreover it is required that no variable in $R'$ occurs in the tree on the
path from the root to $G'$.

\item\label{sls.negation}
Consider a node $G'=C,\oll,\neg A,\olll$ of the t-tree (tu-tree), in which the
selected literal $\neg A$ is negative.  The node is a leaf or has one child,
under the following conditions.
\begin{enumerate}
\item
If the tree is a t-tree then 
\begin{enumerate}
\item\label{sls.negation.t.leaf}
$G'$ is a leaf, or
\item\label{sls.negation.t.child}
$G'$ has a child $C', C,\oll,\olll$, where
$C'$ is a negative answer of a tu-tree for $C,A$ of rank $<k$,
and  $C',C$ is satisfiable.
\end{enumerate}

\item
If the tree is a tu-tree then 
\begin{enumerate}
\item\label{sls.negation.tu.child.trivial}
$G'$ has a child $C,\oll,\olll$, or

\item\label{sls.negation.tu.child}
$G'$ has a child $C', C,\oll,\olll$, where
$C'=\neg C''$  is the negation of an answer $C''$ of a t-tree
for $C,A$ of rank $<k$,
and  $C',C$ is satisfiable,
or
\item\label{sls.negation.tu.leaf}
$G'$ is a leaf and there exists  an answer $C''$ of a t-tree
for $C,A$ of rank $<k$ such that $\neg C'',C$ is unsatisfiable.
\end{enumerate}
\end{enumerate}

\end{enumerate}

\end{df}

An informal explanation for case \ref{negative.answer} is that 
the constraints of the cross-section include all the cases in which $G$ is
\ttt or 
\uuu, thus their negation implies that $G$ is \fff.
A useful intuition is that adding a negative answer $C$ to the nodes of the
tu-tree results in a failed tree -- a tree for $C,G$ without any successful
leaf. 
(For the constraint $C_i$ of any node of the cross-section,
the constraint  $C,C_i$ is unsatisfiable.  The same holds for 
any node which is a descendant of some node of the
cross-section.)

\commenta{ 
}

An informal explanation for case \ref{sls.negation}  is that 
in a t-tree (case \ref{sls.negation.t.child})
$C'$ implies that  $C,A$ is \fff, equivalently 
$\neg (C, A)$ is \ttt.
Hence $C',C$ implies that $\neg A$ is \ttt.
In a tu-tree (\ref{sls.negation.tu.child})
$\neg C''$ includes all the cases in
which $C,A$ is not \ttt.
Hence $\neg C'',C$ -- the constraint of the child --
includes all the cases in which $A$ is not \ttt, 
equivalently in which $\neg A$ is \ttt or \uuu.

\commenta{
}

Notice that in case \ref{sls.negation.t.leaf}
the node $G'=C,\oll,\neg A,\olll$ may
unconditionally be 
a leaf of a t-tree (of any rank).  This corresponds to the fact that
 $C'=\neg C$ is a negative answer for any tu-tree for $C,A$.
(Take the cross-section $\{\, C,A \,\}$).
Hence in the supposed child of $G'$ (case \ref{sls.negation.t.child})
the constraint $\neg C,C$  is unsatisfiable. 
Conversely, according to \ref{sls.negation.tu.child.trivial}, node
$G'=C,\oll,\neg A,\olll$ in a tu-tree may have $C,\oll,\olll$ as the child.
This corresponds to the fact that $C''=\bf false$ is an answer of any t-tree.
Hence $C$ is equivalent to $\neg C'',C$ (which is the constraint obtained in
\ref{sls.negation.tu.child}).
Thus \ref{sls.negation.tu.child.trivial} is a special case of 
\ref{sls.negation.tu.child}.

\begin{example}
\label{ex:opsem}
Consider a query $w(c)$ for the hybrid program of Example~\ref{ex:hyrules}. 
It can be  answered by  the operational semantics by  construction
of the following trees.
(Sometimes we replace a constraint by an equivalent one.)
\begin{enumerate}
\item A t-tree for $w(c)$:
\vspace{-1pt plus 1pt minus 2ex}
\[
\begin{array}{cc}
\multicolumn{2}{c}{   w(c)  } \\
\multicolumn{2}{c}{   |  } \\
\multicolumn{2}{c}{X\eq c, m(X,Y), \neg w(Y) } \\
\multicolumn{2}{c}{ / \qquad\qquad\qquad\qquad \backslash   } \\
X\eq c,Y\eq f,\neg Fi(f),\neg w(Y)  &      X\eq c, Y\eq d, \neg w(Y) \\
 | & | \\
X\eq c,Y\eq f, \neg Fi(f)   &   X\eq c, Y\eq d, \neg(X\eq c, Y\eq d, \neg E(f))
\end{array}
\]
The tree refers to negative answers derived in the cases \ref{case.w.d},
\ref{case.w.f} below. 
The constraint in the second leaf is equivalent to 
$X\eq c, Y\eq d,  E(f)$
(as $\alpha\land\neg(\alpha\land\beta)$ is equivalent to
$\alpha\land\neg\beta$\/).
The answer obtained from the two leaves $C_1,C_2$ of the tree
is  $\exists X,Y\, (C_1\lor C_2)$.
It is equivalent to $\neg Fi(f) \lor E(f)$.
As this constraint is a logical consequence of the ontology,
$w(c)$ holds in each well-founded model of the program. 

\item\label{case.w.d}
\comment{
}
 A tu-tree for $X\eq c,Y\eq d, w(Y)$,
employing an answer from the t-tree from case \ref{case.w.e}:
\[
\renewcommand{\eq}{\eqlooser}
\begin{array}{c}
       X\eq c, Y\eq d, w(Y) \\
| \\
       X\eq c, Y\eq d, X'\eq Y, m(X',Y'), \neg w(Y') \\
| \\
       X\eq c, Y\eq d,          X'\eq d, Y'\eq e, \neg w(Y')   \\
| \\
       X\eq c, Y\eq d, X'\eq d, Y'\eq e,
            \neg(X\eq c, Y\eq d, X'\eq d, Y'\eq e, E(f))
\end{array}
\]
The leaf is equivalent to $X\eq c, Y\eq d, X'\eq d, Y'\eq e, \neg E(f)$,
see the explanation in the previous case.
Hence from the cross-section containing the leaf we obtain a negative answer
equivalent to
$\neg (X\eq c, Y\eq d, \neg E(f))$
 and to $D = \neg (X\eq c, Y\eq d) \lor E(f)$.

Informally, $D$ implies falsity of the root of the tu-tree.
(For a formal treatment see Theorem \ref{theo.soundness} 
and Lemma \ref{lemma.soundness} below).
Hence $E(f)$ implies $\neg w(d)$.
Formally, if $M_0\models E(f)$ for some model $M_0$ of \T
then 
then $w(d)$ is false in $WF(P/M_0)$
(the well-founded model of $P$ based on $M_0$).

\comment{
}

\item\label{case.w.e}
 A t-tree for $Y'=e, w(Y')$ employing a negative answer from case
 \ref{case.w.f}:
\[
\begin{array}{c}
               Y'= e, w(Y')   \\
| \\
               Y'=e, X''=Y',m(X'',Y''),\neg w(Y'')   \\
| \\
              Y'=e, X''=Y', X''=e, Y''=f,E(f),\neg w(Y'')   \\
| \\
               Y'=e, X''=e,Y''=f,E(f) 
\end{array}
\]
The corresponding answer is (equivalent to)  $Y'\eq e,\,E(f)$.
Informally, the answer implies $Y'= e, w(Y')$.
From Lemma \ref{lemma.soundness} below it follows that if
$E(f)$ holds in some model $M_0$ of \T then $w(e)$ is true in the
corresponding well founded model of $P$.

Notice that if $C,Y'\eq e$ is a satisfiable constraint then $C$ may be added to
the nodes of the tree (maybe with renaming of variables $X'',Y''$).
Hence  $C,Y'\eq e,E(f)$ is an answer for $C,Y'\eq e, w(Y')$.
To construct the t-tree of case \ref{case.w.d} we use 
$C = (X\eq c, Y\eq d,          X'\eq d) $.
{\sloppy\par}
\comment{
}

\item\label{case.w.f} A tu-tree for  $Y\eq f, w(Y)$,
with atom $m(X',Y')$ selected in the leaf:
\[
\begin{array}{c}
                   Y\eq f,w(Y)         \\
| \\
       Y\eq f, Y\eq X',m(X',Y'),\neg w(Y') \\
\end{array}
\]
 From the empty cross-section a negative answer
{\bf true} is obtained.
So $w(f)$ is false in the well-founded semantics of the program.
Similarly, {\bf true} is a negative answer for $C,\, Y\eq f,\, w(Y)$,
where $C$ is an arbitrary constraint.
\end{enumerate}
\end{example}

Various simplifications of t- \mbox{(tu-)} trees are possible.  For instance
in case \ref{case.w.e} of the last example
the nodes of the tree may be replaced by
$w(e);\ \linebreak[3] m(e,Y), \neg w(Y);\
 E(f),\neg w(f);\ E(f)$.
This issue is outside of the scope of this paper. 

We do not deal here with
actual implementing of the operational semantics.  (An implementation is
described in \cite{DrabentHM07}.)
We only mention that -- similarly as in CLP -- it is not necessary to
check satisfiability of the constraint for each node.
The answers (negative answers) of trees obtained in this way are
logically equivalent to those of t-trees (tu-trees) from
Def.\ \ref{def.operational}.

\comment{
}
\comment{
}

\subsection{Soundness}\label{sec:soundness}
\comment{
}

\comment{
}

In this section we prove soundness of the operational semantics of hybrid
programs (Def.\,\ref{def.operational}) with respect to their declarative
semantics (Def.\,\ref{def.declarative}).
Before the actual proof we discuss ground instances of goals and trees,
and introduce safe programs and goals.
These notions are employed in the proof.

\comment{
}

For our proofs we use the characterization of the well-founded semantics
of logic programs from Section~\ref{sec:wfs}.
So
for a given model $M_0$ of the external theory,
the well-founded model of the program is 
$\psipmpower\alpha$ for some $\alpha$.

\subsubsection{Ground instances of trees}

By an {\bf extension} of a substitution $\theta$ we mean any
substitution of the form  $\theta\cup \theta'$
(where $\theta=\{x_1/t_1,\ldots,x_n/t_n\}$
 and $\theta'=\{y_1/u_1,\ldots,y_n/u_n\}$ are substitutions with
disjoint domains,
$\{\seq x\}\cap\{\seq[m]y\}=\emptyset$).

By a {\bf grounding substitution} for the variables of a formula $F$ 
(or just ``for $F$'')
we mean a substitution
replacing the free variables of $F$ by ground terms.
(The domain of the substitution may include other variables.)

Let $G=C,\oll$ be a goal and $M_0$ a model of \T.
  Let $\theta$ be a grounding substitution for the variables of $G$.
(Notice that $C\theta$ has no free variables.)
If $C\theta$ is true in $M_0$ then
we say that $\theta$ is {\bf applicable} to $G$ (w.r.t.\ $M_0$), and
by the result $G\theta$ of applying $\theta$ to $G$
we mean the ground goal $\oll\theta$;
it is called a {\bf ground instance} of $G$.
Similarly, we say that $\theta$ is {\em applicable} to a rule
$H\gets C,\oll$; the result $H\theta\gets\oll\theta$
is called a {\bf normal ground instance} of the rule.%
\footnote{
  Definition \ref{def.declarative} employs another kind of
  ground instance, namely  $H\theta\gets C\theta,\oll\theta$.
  As ground instances in that sense are
  not used below, we sometimes skip the word ``normal.''
  To simplify notation, we write $R\theta$ for a normal ground
  instance of a rule $R$, when this does not lead to ambiguity.
}
Notice that a rule $R\theta$ is a normal ground instance of a rule
$R\in P$ w.r.t.\ $M_0$ iff $R\theta\in (P/M_0)$.

Consider a t-tree or tu-tree $\Tr$ 
for $G$ and $\theta$ as above.
A {\bf ground instance} $\Tr\theta$ of $\Tr$ w.r.t.\ $M_0$ is defined
recursively as follows. 
The nodes of $\Tr\theta$ are ground instances of (some) nodes of $\Tr$,
each node $H$ of $\Tr\theta$ {\bf corresponds} to a node $G'$ of $\Tr$
such that $H$ is a ground instance of $G'$.
The root of  $\Tr\theta$ is $G\theta$ and it corresponds to the root $G$ of $\Tr$.
If a node $G'\theta'$  of  $\Tr\theta$ corresponds to node $G'$ of $\Tr$
(where $\theta'$ is a grounding substitution for the variables of $G'$),
 $G''$ is a child of $G'$ in $\Tr$, $\theta''$ is an extension of $\theta'$
then  $G''\theta''$ is a child of $G'\theta'$ in $\Tr\theta$,
provided that $\theta''$ is applicable to $G''$.

A node of $\Tr\theta$ corresponding to a successful leaf of $\Tr$
will be called a {\bf successful} leaf of  $\Tr\theta$.

\comment{
}

\begin{example}
Consider a program $P=\{p(a)\gets q(y)\}$.
The t-tree (or tu-tree) $\Tr$ for $p(x)$ consists of two nodes;
the child of $p(x)$ is $x=a,q(y)$.
For $\theta=\{x/b\}$ the ground instance $\Tr\theta$ consists of one node
$p(b)$.
For $\sigma=\{x/a\}$  the root $p(a)$ of $\Tr\sigma$
has a child $q(t)$ for each ground term $t$.
Each node $q(t)$ in  $\Tr\sigma$ corresponds to the node 
$x=a,q(y)$ of \Tr.
\end{example}

\comment{
}

Notice that if $G'\theta'$ and its child $G''\theta''$ in $\Tr\theta$
 correspond, respectively, to
$G'$ and $G''$ in \Tr, and $G''$ is derived from $G'$ by a rule
$R\in P$ 
then  $G''\theta''$ is derived from  $G'\theta'$ by a normal ground instance
$R\sigma\in (P/M_0)$ of $R$.
(This means that $R\sigma$ is $H\mathop\gets \ol B$, $H$ is an atom in
 $G'\theta'$  and $G''$ is  $G'\theta'$ with $H$ replaced by $\ol B$.)
Thus if a leaf  $G'\theta'$ of $\Tr\theta$
corresponds to a node  $G'$ of $\Tr$,
and a positive literal $A$ is selected in $G'$
then no normal ground instance of a rule in the program has the head
$A\theta$.
If no negative literal is selected in \Tr  then $\Tr\theta$ is an SLD-tree
(\cite{lloyd87foundation}) for program $(P/M_0)$.

\subsubsection{Safeness}

We introduce a notion similar to DL-safeness \cite{MotikSS05,Rosati06},
but taking into account that constraints may contain equality $=$.

\begin{df}
Let $C$ be a constraint. 
A variable $x$ is {\bf bound} in $C$ to a ground term $t$ (to a variable $y$)
if  $\T \models C \to x=t$
(respectively $\T \models C \to x=y$).
\end{df}

For instance, in $x\mathbin=f(y),\,y\mathbin=a$
variable $x$ is bound to ground term $f(a)$,
and $y$ is bound to $a$.
Notice that any variable is bound to itself independently of the constraint.
A simple sufficient condition is that,
for variables $x_0,\ldots,x_n$  ($n\geq0$) and a ground term $t$,
 if 
$C$ is a conjunction of
constraints $\seq[l] C$ and 
 the set  $\{\seq[l] C\}$ contains equalities
$x_0\eq x_1,\ldots,\linebreak[3]x_{n-1}\eq x_n$ 
(resp.\ 
$x_0\mathop=x_1,\ldots,\linebreak[3] x_{n-1}\eq x_n,x_n\eq t $)
then $x_0$ is bound to $x_n$
(resp.\ to~$t$) in $C$.

\begin{df}
A rule
 $R=H\gets C,\oll$, where $C$ is the constraint of $R$,
is {\bf safe} if 

 -- each variable of $H$,

 -- each variable of a negative literal of \oll, and

 -- each free variable of $C$

\noindent
is bound  in $C$ to  a ground term or to a variable appearing
 in a positive literal in~\oll.

Let $V$ be a set of variables.
When the conditions above are satisfied with possible exception for the
variables from $V$ then we say that $R$ is {\bf safe apart} of $V$.

A set of rules is {\em safe} if all its rules are safe.
A hybrid program \hp is is {\em safe} if $P$ is safe.
A goal $G=C,\oll$ is {\em safe} if the rule
$p\gets G$ is safe (where $p$ is a 0-argument predicate symbol).
$G$ is {\em safe apart} of $V$ if the rule
$p\gets G$ is safe apart of $V$.

\end{df}

If the root of a t-tree (tu-tree) for a safe program is safe then any node of
the tree is safe. 
Hence, in the constraint of a successful leaf, all the free variables are bound
to ground terms.
In the Appendix we prove a more general property:

\begin{lemma}
\label{lemma.safe}

Let \hp be a safe program, $V$ a set of variables, and $G$ a goal.
 Consider a t-tree (or a tu-tree) with the root $G$.

1.\mbox{\ \ }%
Each node of the tree is safe apart of $V_0$,
where $V_0$ is the set of free variables of $G$.

2.\mbox{\ \ }%
Assume that no variable from $V$ occurs in any
variant of a rule from $P$ used in constructing the tree.
If $G$ is safe apart of $V$ 
then each node of the tree is safe apart of $V$.

\end{lemma}

\subsubsection{Soundness theorem}

We will say that the set of constraints has the {\bf witness property}
if for any model $M$ of the external theory \T
and for any constraint $C$,
whenever $M\models \exists C$ then 
$M\models C\theta$ for some grounding substitution $\theta$ for $C$. 
The witness property is implied by the parameter names assumption (PNA)
\cite{BruijnPPV07}
that restricts the interpretations of \T to those in which every domain
element is a value of a ground term. 

The operational semantics may be not sound for constraints without witness
property and non safe programs.  As an example take
$P=\{\, p\gets q(x) \,\}$
 and $\T = \{\exists x.q(x)\}$.  
Then (a constraint $\exists x.q(x)$ equivalent to) {\bf true}
is an answer of the t-tree for $p$.
However $\hp\notmodels_{\mathrm{wf}}p$
(as there exist models of \T in which every ground instance of $q(x)$ is
false).

\begin{lemma}[Soundness]\label{lemma.soundness}
Consider a program \hp,
a goal $G$,  a model $M_0$ of~{\T}, and a countable ordinal
number $k$.
Assume that $M_0$ is a Herbrand interpretation, or the set of constraints
has the witness property, or $P$ is safe.

1.\mbox{\ \ }%
If $C$ is an answer of a t-tree of rank $k$ for $G$ then
for any grounding substitution
$\theta$ (for the variables of $G$)
$M_0\models C\theta$ implies $\psipmpower {k+1}\models_3 G\theta$.

2.\mbox{\ \ }%
If $C$ is a negative answer of a tu-tree of rank $k$ for $G$ then
for any grounding substitution
$\theta$ (for the variables of $G$)
$M_0\models C\theta$ implies $\psipmpower {k+1}\models_3 \neg G\theta$.
\sloppy
\end{lemma}

The proof is presented in the Appendix.%
\footnote{
   From the proof it follows that safeness is needed only for the rules
   that have been used in constructing t-trees
   (the t-trees referred to, directly or indirectly, by the t- (tu-) tree for
   $G$). 
   Alternatively,
   the witness property is necessary only for the constraints that are
   successful leaves of these t-trees.
}
It is based on studying ground instances of the t- (tu-) tree for $G$
and viewing them as SLD-trees of a program \PMtI 
(respectively a program related to \PMtuI)
from the definition of~\psipm.

\commenta{
}

\medskip

As a corollary we obtain:

\comment{
}

\begin{theorem}[Soundness]
\label{theo.soundness}
Let \hp be a hybrid program and $G=C_0,\oll$ a goal
(where $C_0$ is the constraint of $G$).
Assume that  $P$ is safe, or the set of constraints has the witness property.

If $C$ is an answer of a t-tree for \hp and  $G$ then,
for any substitution~$\theta$,
$\T\models C\theta$ implies $\hp\models_{\rm wf} \oll\theta$.
  
If $C$ is a negative answer of a tu-tree for \hp and $G$ then,
for any  substitution
$\theta$,
$\T\models C\theta$ implies \mbox{$\hp\models_{\rm wf} \neg \oll\theta$}.
\end{theorem}

\comment{
}
\noindent
PROOF. \  Let $G=C_0,\oll$ and $C$ be an answer of the t-tree.
$\T\models C\theta$ implies $M_0\models C\theta\theta'$ for any
substitution $\theta'$ and any model $M_0$ of \T.
Consider a $\theta'$ such that $x\theta'$ is ground for each free variable
$x$ of $(C,\oll)\theta$.
From Lemma \ref{lemma.soundness}, applied to $M_0$, $C$, $G$, and the
substitution $\theta\theta'$,
it follows that
 $M\models_{3} G\theta\theta'$,
where $M$ is the well-founded model or $P$ based on $M_0$.
Notice that $G\theta\theta'= \oll\theta\theta'$.
As $M\models_3\oll\theta\theta'$ for each  $\theta'$ as above
and $M$ is a Herbrand interpretation,
we have $M\models_3\oll\theta$.
As the latter holds for each well-founded model of $P$,
we obtain $\hp\models_{\rm wf} \oll\theta$.

The proof for a negative answer of a tu-tree is analogical. \hfill $\Box$

\comment{
}%

\medskip

It may be desirable to have an operational semantics which is sound
also for non safe programs and constraints without the witness
property.
This can be obtained by employing, in formula restrictions
(Def.\ \ref{def.restriction...}),
 a certain non standard quantifier 
$\exists'$ instead of $\exists$. 
For the new quantifier it holds that
if $I \models \exists' F$ then 
$I \models F\theta$ for some grounding substitution $\theta$ for $F$
(for any formula $F$ and interpretation $I$).
The details are outside the scope of this paper.

\subsection{Completeness}\label{sec:completeness}

In a general case our operational semantics is not complete.
Roughly speaking, the reason is using only finite constraint formulae 
as (negative) answers%
\footnote{
In \cite{drabent93-lpnmr,drabent95}
this problem was solved by allowing an infinite set of children of a
node with a negative literal selected.
Example 4.10 in \cite{drabent93-lpnmr} shows that this is
actually necessary.  Thus it provides a counterexample for
completeness of the operational semantics of Def.\ \ref{def.operational}.
}
in case \ref{sls.negation} of Def.\ \ref{def.operational}.
We show completeness of our operational semantics for the case where
the Herbrand universe is finite and 
the program and goals are safe.
The completeness result includes independence from the selection rule.

We first present a
technical lemma about
simplifying goals for which t-trees (tu-trees) are constructed.
Then we restrict our considerations to safe programs over a finite
universe, 
show how a kind of a most general (negative) answer for a given
tree can be obtained, and define a notion of a maximal t- (tu-) tree.
Intuitively, a maximal tree (of a sufficiently high rank) derives
everything that is required by the declarative semantics.
This is made formal in a completeness lemma, from which
completeness of the operational semantics follows.

\comment{
}

The following lemma shows how a tree for $A$ may replace a
tree for $C,A$.  So, for a fixed $A$, many trees for goals
$C,A$ may be replaced by a single tree.  (On the other hand, for a
fixed $C$ the tree
for $A$ may have more nodes than the corresponding one for  $C,A$.)

\begin{lemma}\label{lemma.A.CA}
Consider a program \hp.
 Let $A$ be an atom and $C$ a constraint.
If $\neg(\restrict{C_1}A),\ldots,\neg(\restrict{C_n}A)$
is a negative answer of a tu-tree
(respectively a negation of an answer of a t-tree)
of rank $k$ for $A$ then
$\neg\bigl(C,(\restrict{C_1}A)\bigr),\ldots,
\neg\bigl(C,(\restrict{C_n}A)\bigr)$,
or equivalently
$\neg C \lor\neg(\restrict{C_1}A),\ldots,\neg(\restrict{C_n}A)$,
is a negative answer of a tu-tree (the negation of an answer of a t-tree)
of rank $k$ for $C,A$.
{\sloppy\par}
\end{lemma}

\noindent
PROOF. \
Without loss of generality we can assume that if a variable $x$ occurs
both in the tree for A and in C then $x$ occurs in A.
If $C',\ol L$ is a node of the tree for $A$ then
$C,C',\ol L$ is a node of the tree for $C,A$ provided that $C,C'$ is
satisfiable. 
Assume the negative answer for $A$ is obtained from a finite cross-section.
Consider a ``corresponding''
 cross-section of the tree for $C,A$ whose node constraints
are those of $C,C_i$ that are satisfiable.
The corresponding negative answer is 
$\neg(\restrict{(C,C_1)}A),\ldots,\neg(\restrict{(C,C_n)}A)$.
Each
$\restrict{(C,C_i)}{C,A}$ is equivalent to $C,(\restrict{C_i}{A})$.
The cases of an infinite cross-section, and of a t-tree, are similar.
\hfill ${\Box}$

\subsubsection{Maximal trees}

For this section we assume that the Herbrand universe is
finite.  Hence
the alphabet of function symbols is finite and contains only constants.

A t- (tu-) tree may be infinite.  Moreover, the set of nodes in the
tree with a negative literal selected may be infinite.  Hence the tree
may refer to an infinite set of subsidiary trees.  The answers
(negative answers) of the tree may be obtained from an infinite set of
successful leaves (an infinite cross-section); so it seems that 
we have to deal with an infinite set of answers (as there 
may not exists one which implies all the others). 
In what follows we show how to
avoid infinite sets of answers.

Notice first that 
the set of selected negative literals in a t- (tu-) tree is finite,
up to renaming of variables (as the Herbrand universe is finite).
By Lemma \ref{lemma.A.CA}, instead of constructing a possibly infinite
set of subsidiary tu- (t-) trees for goals of the form $C,A$,
it is sufficient to construct a finite set of tu- (t-) trees for goals
of the form $A$.
\comment{
}

 In a t- (tu-) tree with a safe root, if $C$ is a successful leaf
then each free variable $x$ of $C$ is bound to a constant $c_x$.
This defines a grounding substitution
$\theta=\{x_1/c_{x_1},\ldots,x_m/c_{x_m}\}$
for the free variables $\seq[m]x$ of $C$.
Now $C$ is equivalent to $x_1=c_{x_1},\ldots,x_m=c_{x_m},C$
and to $x_1=c_{x_1},\linebreak[3]\ldots,\linebreak[3]x_m=c_{x_m},C\theta$.

By a {\bf grounded constraint} from a program
 $P$ (from a goal $G$) we mean a
constraint 
$C\sigma$ such that constraint $C$ is the constraint of a rule of $P$
(the constraint of $G$) and the substitution $\sigma$ replaces
all the free variables of $C$ by constants.
A constraint is in a {\bf solved form} for a program $P$ and an initial
goal $G$ if it is
a conjunction of constraints of the form $x=c$, $C_0$ or $\neg C_0$,
where $x$ is a variable, $c$ is a constant, and $C_0$ is a grounded
constraint from $P$ or $G$.
 The set of grounded constraints from a given program
or a given goal is finite.
So the set of constrains in solved form (for $P$ and $G$) with 
the free variables from $G$ 
is finite, up to equivalence.

We now show that, under certain conditions,
the successful leaves of t- (tu-) trees
may be seen as disjunctions of constraints in solved form.

\begin{lemma}\label{lemma.solved.form}
Let the Herbrand universe be finite and
 $\Tr$ be a t-tree (tu-tree) for a safe goal $G$ and a safe program \hp.
Assume that each negative answer (negation of an answer)
employed in the tree is of the form $\seq{\neg C}$,
where each $C_i$ is in solved form for $P$ and $G$.
Then the constraint of any success leaf $C$ of the tree
is equivalent to a disjunction of constraints in solved form.
Also, $\restrict C G$ 
is equivalent to a disjunction of constraints in solved form for $P$ and $G$.
\end{lemma}
\noindent
PROOF. \ 
Consider a success leaf $C$.
Each free variable $x$ of $C$ is bound in $C$ to a constant $c_x$.
Let $\theta=\{x_1/c_{x_1},\ldots,x_m/c_{x_m}\}$ (where $\seq[m]x$
are the free variables of $C$).
$C$ is equivalent to $x_1=c_{x_1},\ldots,x_m=c_{x_m},C$
and to $x_1=c_{x_1},\ldots,x_m=c_{x_m},C\theta$
($C\theta$ has no free variables).

$C\theta=\seq{C}$, where
each $C_i$ is a ground equality, or a grounded constraint from $P$ or
from $G$, 
or a $\neg C'_i\theta$ where $C'_i$ is a constraint in a solved form.
$\neg C'_i$ is equivalent to $\neg D_1\lor\ldots\lor\neg D_l$, where each
$D_i$ is of the form $x=c$ or is a possibly negated grounded constraint
from $P$ or from  $G$.
Thus $\neg C'_i\theta$ is equivalent to
$\bf true$ (if some ground disequality $\neg D_i\theta$ is true), or to 
$\neg D_{j_1}\lor\ldots\lor\neg D_{j_{l'}}$, where each $D_{j_i}$
is a possibly negated grounded constraint from $P$ or $G$.
Each $C_i$ which is a ground equality is equivalent to $\bf true$
(as it is satisfiable).

So applying distributivity
 ($\phi,(\psi\lor\psi')\equiv\phi,\psi\lor\phi,\psi'$) to
 $x_1=c_{x_1},\ldots,\linebreak[3]x_m=c_{x_m},\linebreak[3]C\theta$
 results in an equivalent disjunction of constraints in solved form.
Removing from the latter constraint each $x_i=c_i$ where $x_i$ not free
in $G$ produces a disjunction of constraints in solved form equivalent to 
$\restrict C G$. \hfill $\Box$
\medskip

As discussed above, 
the set of disjunctions of constraints in solved form (for $P$ and $G$) with 
the free variables from $G$ is finite, up to equivalence.  Thus
 from the lemma it follows that the set of success leaves of the t-tree
(tu-tree) is equivalent to a finite set of disjunctions of constraints
in solved form.
Formally: 
If the tree satisfies the conditions of Lemma \ref{lemma.solved.form} then
there exists a constraint $D$ (we will call it a {\bf finite answer} of the
t-tree,
resp.\ {\bf finite pseudo-answer} of the tu-tree),
such that
\[
\mbox{
 $M_0\models \restrict C G\theta$ for some success leaf $C$ of the
tree \quad iff \quad $M_0\models D\theta$,
}
\]
for every model $M_0$ of \T and
 every substitution $\theta$ grounding the variables of $G$.
\comment{
}%
Moreover, $D$ is a disjunction of constraints in solved form,
for $P$ and the root $G$ of the tree.
($D$ is equivalent to a constraint
$(\restrict{C_1}G)\lor\ldots\lor(\restrict{C_m}G)$
where each $C_i$ is a success leaf of the tree.)

\comment{
}

Notice that the negation of a pseudo-answer of a tu-tree is a negative
answer of the tree.
(The corresponding cross-section contains all the successful leaves.)
Informally, a finite answer is a most general answer that can be obtained from
a given t-tree, and 
the negation of a finite pseudo-answer is a most general negative answer that
can be obtained from a given tu-tree.

Now for a given rank and goal we define a maximal t- (tu-) tree.
The intention is that the tree provides a (negative) answer
which is more general than other (negative) answers for this goal of
the same rank. 

\begin{df}\label{def.maximal.tree}
Let the Herbrand universe be finite,
and \hp be a safe program.
A {\bf maximal} t-tree (tu-tree) for a goal and \hp is defined
inductively: 

A {maximal} t-tree of rank $0$
is a t-tree in which each node with a negative literal selected is a leaf.
  
A {maximal} tu-tree of rank $0$ is a tu-tree
in which if a negative literal is selected in a node $G'$
and the constraint of $G'$ is $C$ then the constraint of the child 
of $G'$ is $C$
(cf.\ Def.\ \ref{def.operational} case \ref{sls.negation.tu.child.trivial}).

A maximal t-tree (tu-tree)
of rank $k>0$ is a t-tree (tu-tree) in which
a node $G'$ with a negative literal $\neg A$ selected has a child $G''$
iff $C,\neg D$ is satisfiable, where
$C$ is the constraint of $G'$ and $D$ is 
a finite pseudo-answer (a finite answer) of a maximal tu-tree (t-tree)
for $A$ of rank $k-1$;
moreover  $C,\neg D$ is the constraint of  $G''$.%
\footnote{
By Lemma \ref{lemma.A.CA},
$\neg C\lor\neg D$ is a negative answer (the negation of an answer)
for $C,A$.  Hence according to Def.\ \ref{def.operational},
the constraint of the child of $G'$ is $C,(\neg C\lor\neg D)$, which is
equivalent to $C,\neg D$.
\comment{
}%
}%

\comment{
}
\end{df}
The definition is correct:
  From Lemma \ref{lemma.A.CA} and \ref{lemma.solved.form}
by induction on the rank we obtain that,
for a safe program $P$ and a safe goal $G$,
the tree described in the definition
satisfies the conditions of Lemma \ref{lemma.solved.form},
it has a finite \mbox{(pseudo-)} answer which is a disjunction of constraints
in solved form;%
\footnote{
For $P$ and $G$.
If  $G$ consists of a single atom $A$ then the constraints are in
solved form for $P$ and any goal.
}
hence the finite (pseudo-) answers employed in the definition exist.
Notice that a maximal t- (tu-) tree is defined for any safe program
and any goal $G_0$.
If $G_0$ is safe then the tree has a finite (pseudo) answer.

\commenta{
}

\subsubsection{Completeness theorem}
By a {\bf selection rule} we mean a function which,
given a sequence of goals $G_0,\ldots,G_n$,
selects a rule literal in $G_n$
provided $G_n$ contains at least one rule literal
and $G_0,\ldots,G_n$ is a prefix of a branch of a t-, tu-, or SLD-tree.

\comment{
}

Now we are ready to state and prove completeness of the operational
semantics. 
We first show that a (negative) answer of any maximal (tu-) t-tree for
$G$ of rank $j$ 
-- speaking informally --
describes all the instances of $G$ that are true (false) in
a corresponding 
approximation $\psipmpower{k}$ of the well-founded model of $P/M_0$.

\begin{lemma}[Completeness]
\label{lemma.completeness}
Assume that the Herbrand universe is finite.
Consider a safe program \hp.
Let $G$ be a safe goal, $C_0$ be the
constraint of $G$, and $k>0$ be a natural number.
Consider a selection rule $\cal R$, and a maximal t-tree and a maximal tu-tree
for $G$ of rank $k-1$ if $G$ does not
contain a negative literal, and of rank $k$ otherwise.

Let $M_0$ be a model of $\T$, and
$\theta$ be a grounding substitution for the variables of $G$ such that
$M_0\models C_0\theta$
\comment{
}

1.\mbox{ \ }If $\psipmpower k\models_3 G\theta$
and $C$ is a finite answer of the t-tree
then $M_0\models C\theta$.

2.\mbox{ \ }If $\psipmpower k\models_3\neg G\theta$
and $D$ is a finite pseudo-answer of the tu-tree
(and thus $\neg D$ is a negative answer)
then $M_0\models\neg D\theta$.
\comment{
}
\comment{
}

\end{lemma}

The basic idea of the proof is that in case 1.\ it follows that
there exists a successful SLD-derivation for $G\theta$ and a certain program
\PMt J.  This derivation is shown to be a branch of a ground instance
of the maximal \mbox{t-tree} for $G$. 
In case 2.\ there does not
exist a successful SLD-derivation for $G\theta$ and a certain program
\PMtu J.
However such a derivation is shown to be a branch of a ground instance of the
maximal tu-tree for $G$, under assumption that 
$M_0\models D\theta$.  Hence  $M_0\models\neg D\theta$.
The detailed proof is given in the Appendix.

  \commenta{
}

\medskip

As a main result we obtain completeness and independence
from the selection rule (of the operational semantics 
of Def.\ \ref{def.operational}
w.r.t.\ the declarative semantics of Def.\ \ref{def.declarative}).

\begin{theorem}[Completeness]
Assume that the Herbrand universe is finite.
Consider a safe program $\hp$, a safe goal $G=C_0,\oll$
(where $C_0$ is the constraint of $G$), and a selection rule \R.
Let
 $\theta$ be a grounding substitution for the variables of $G$ such that
 $C_0\theta$ is  satisfiable.

1. If $\hp \models_{\rm wf} \oll\theta$
then there exists a t-tree (of a finite rank)
for $G$ via $\R$ with an answer $C$
such that $\T\models C\theta$.

2. If $\hp \models_{\rm wf}\neg\oll\theta$
then there exists a tu-tree (of a finite rank) for $G$ via $\R$
with a negative answer $C$
such that $\T\models C\theta$.
\end{theorem}

\noindent\pagebreak[3]%
PROOF. \ 
As the Herbrand universe is finite, 
the set of possible programs $P/M_0$ is finite.
Thus there exists a natural number $k$ such that $\psipmpower k$ is the
well founded model of $P/M_0$ for each $M_0$.
Now the Theorem follows from Lemma \ref{lemma.completeness}.
\hfill
$\Box$

\section{Decidability}\label{sec:decidable}

Now we show that the well-founded semantics for hybrid programs is decidable
in the case of Datalog, i.e.\ when the set \F of function symbols is a finite
set of constants.  (No safeness condition is needed.)
The proof employs the soundness and completeness results from the previous
sections. 

\begin{theorem}[Decidability]
\label{theo.decidability}
Assume that the Herbrand universe is finite, and that
for any closed constraint $C$ it is decidable whether $\T\models C$.
There exists an algorithm which, for a hybrid program \hp
and a ground atom~$A$, finds out whether  $\hp \models_{\rm wf} A$
and whether $\hp \models_{\rm wf}\neg A$.
\end{theorem}

\newcommand*{\hpground}{\ensuremath{(ground(P),{\cal T})}\xspace}

\noindent PROOF. \
We first show that each maximal t- and tu-tree for 
program \hpground can be represented and constructed in a finite way.

We say that a constraint is in {\em ground solved form} if it is 
built out of the constraints of the rules of $ground(P)$
by means of $\neg$ and $\land$.
The set of such constraints up to logical equivalence is finite.
(Convert them to a disjunctive normal form, remove repeated literals in the
conjunctions, remove repeated conjunctions.)

Let $B$ be a ground rule atom, and 
\Tr be a t-tree or a tu-tree for $B$.
Each satisfiable equality constraint that appears in the tree
is of the form $a=a$ (where $a\in\F$), hence it is valid and may be removed.
Assume that each lower rank (negative) answer employed in the
tree is in ground solved form.
Then each constraint in the tree is in ground solved form.
So the set of these constraints is finite up to equivalence.

The set of rule literals that appear in \Tr is a subset of the literals of
$ground(P)$.  Thus the set is finite.
So the set of conjunctions of such literals is finite, up to logical
equivalence. (Repeated occurrences of a literal can be removed.)
As a result we obtain that
the set of nodes of \Tr is finite up to logical equivalence.
\comment{
}

Assume now that the (negative) answers from lower rank trees employed in \Tr
are known.
Then
a finite {\em representation} of \Tr
can be constructed top-down starting from the root $B$.
Before adding a node $G$ to the current
(sub-graph of the) tree, it is checked whether a node $G'$ logically equivalent
to $G$ already exists.  If it does and the same literal is selected in
both nodes then $G$ is not added.
This process terminates,
its result is a tree $\Tr'$ which is a finite sub-graph of \Tr.
Each successful leaf of \Tr is (logically equivalent to) a successful leaf of
$\Tr'$.
In the terminology of the previous section,
the disjunction of the successful leaves of $\Tr'$ is 
a finite (pseudo-) answer of $\Tr$
(and is logically equivalent to each finite (pseudo-) answer of $\Tr$).
In this way finite representations of
the maximal t- and tu-trees for all the atoms can be
constructed, first for rank 0 and then stepwise for ranks $1,2,\ldots\,$.
The process is terminated at rank $k$ when for each atom $A$ the obtained 
finite (pseudo-) answers of rank $k-1$ and  $k$ are equivalent
(as it is then a finite (pseudo-) answer for $A$ of any rank $>k$).

\medskip
The well-founded models of $P$ and those of $ground(P)$ are the same;
hence the programs are equivalent:
$\hp\models_{\rm wf} F$ iff  $\hpground\models_{\rm wf} F$ for any formula
$F$. 
As $ground(P)$ is finite and safe,  the completeness lemma
(\ref{lemma.completeness}) applies.
Thus, for any selection rule~$\R$,
\begin{itemize}
\item  
if
$\hp \models_{\rm wf} A$ \  then \
 $\T\models C$,
for some finite $k\geq 0$
and any maximal t-tree of rank $\geq k$  for $A$ and  \hpground,
with a finite answer $C$,

\vspace{0pt minus 3pt}

\item
if
$\hp \models_{\rm wf}\neg A$ \ then \
$\T\models\neg D$, 
for some finite $k\geq 0$
and any maximal tu-tree of rank $\geq k$ for $A$ and  \hpground,
with a finite pseudo-\linebreak[3]answer $D$.

\end{itemize}
Hence, by the soundness theorem (\ref{theo.soundness}),
$\hp \models_{\rm wf} A$ \  iff \ $\T\models C$
(respectively 
$\hp \models_{\rm wf}\neg A$ \ iff \ $\T\models\neg D$)
for the finite answer $C$ (pseudo-answer $D$) for $A$ computed above.
Checking whether
$\T\models\neg C$  and $\T\models\neg D$ is decidable by the
assumptions of the theorem.
\hfill$\Box$
\medskip

The decidability result should be compared with the fact that
under the assumption of Th.\ \ref{theo.decidability}
it is undecidable whether $P \cup \T \models A$
\cite{levy98carin}.
The difference is that the logical consequence deals with arbitrary
interpretation domains, while the semantics of hybrid programs 
interprets $P$ over the Herbrand universe, which in this case is finite.

\section{Related Work}
\label{sec:disc}

The notions of external theory and constraints used
in hybrid rules are similar to those used in CLP.
However,
classical CLP does not support non-monotonic reasoning.
The results presented in this paper can thus also be seen
as an approach to integration  of both paradigms, based
on the well-founded semantics of normal programs.

We are aware of few papers on non monotonic negation for CLP.
The approach of \cite{Fages97} employs the 3-valued completion semantics.
It is similar to the work presented here;
in both cases the operational
semantics follows the idea of \cite{drabent95}
of selecting a cross section of a derivation tree and negating the
disjunction of respective constraints.
The completion semantics is also used in \cite{Stuckey95}.
An approach generalizing the well-founded semantics for CLP is presented in
\cite{DixS98}.   It however assigns the semantics only to some programs
(those that can be transformed to an irreducible program),
while our semantics deals with all programs.
In contrast to our approach, 
the operational semantics of \cite{DixS98} is not top-down and goal driven.
In consists of applying program transformations to obtain an irreducible
program; the latter can be used to obtain answers to goals.

The presented framework makes it possible to integrate
normal logic programs with ontologies expressed as first-order theories,
including those specified in the  web ontology languages OWL-DL and OWL-Lite.

The problem of integration of rules and ontologies has been addressed
in different ways.
One line of research aims at  achieving  the integration by embedding
rules, ontologies and their combinations in a known logic.
A well-known proposal
of this kind is  SWRL~\cite{SWRL04}, extending ontologies with Horn formulae
within FOL, but not allowing non-monotonic rules. More recent
 attempts~\cite{BruijnET08,MotikR07,BruijnPPV07} address the 
issue of non-monotonicity by embedding non-monotonic rules 
and DL ontologies in various  logics which make it possible to
capture non-monotonicity.

In contrast to that, we achieve the integration by a direct definition
of the semantics of hybrid rules. The declarative semantics combines
the FOL semantics of the external theories with the well-founded
semantics of logic programs.
Negation in the  constraints  of the external theory is
 interpreted in the classical way, while negation in  rule literals
is non-monotonic. We now compare our work with the approaches
to integration of rules and ontologies which make similar
assumptions. We note first that all related work of this kind
is based on Datalog rules, while our approach admits non-nullary
function symbols. 

Our work is strongly motivated by  the early $\mathcal{AL}$-log 
approach \cite{donini98al-log} where  positive  Datalog was
extended by allowing the concepts of 
$\mathcal{ALC}$ DL as constraints in safe Datalog rules.
The operational semantics of $\mathcal{AL}$-log  relies 
on an extension of  SLD-resolution 
where the disjunction of constraints from different 
derivations is to be submitted for validity check to the DL-reasoner.
We adopted the   $\mathcal{AL}$-log idea of extending rules with 
constraints in the body, and applied it to more expressive rules 
including non-monotonic negation,
and to arbitrary external theories of FOL.

In our approach the heads of the hybrid  rules are atoms built with 
rule predicates. Thus the semantics of the rule predicates
depends on the external theory which is assumed to be given
a priori and  not to depend on the rules. The rationale for that
is that the rules describe a specific application while 
the theory (for example an ontology) provides a knowledge common 
for an application domain. In contrast to that, several
papers \cite{MotikSS05,Rosati05,Rosati06} allow the use of ontology 
predicates in the heads of 
rules, defining thus an integrated language where rule predicates
and ontology predicates may be mutually dependent, and ontology
predicates can be \mbox{(re-)}\,defined by rules. 

The paper  \cite{MotikSS05} defines {\em DL rules}, a decidable combination 
of OWL-DL with disjunctive Datalog without non-monotonic negation.
In contrast to that, our primary concern is non-monotonic reasoning. 

The  r-hybrid knowledge bases \cite{Rosati05} and the more recent
${\cal DL}${+}{\it log}
\cite{Rosati06} are based on disjunctive Datalog
with non-monotonic negation under  the stable model semantics.
The objective is to define a generic integration scheme of this variant
of Datalog with an  arbitrary Description Logic. The DL-rules defined
under this scheme may include DL predicates not only in their bodies
but also in the heads. A hybrid ${\cal DL}${+}{\it log} knowledge base consists of 
a DL knowledge base $\cal K$  and a set of hybrid rules \P. 
A notion of {\em non-monotonic  model} of such a knowledge base is defined
by referring to first-order models%
\footnote{
However, only a fixed domain of interpretation is considered, 
with a fixed interpretation of constants.  Moreover,
the interpretation is a bijection
(each domain element is denoted by a distinct constant),
and the domain is countably infinite.
\comment{
}
} %
of $\cal K$
and to the stable models of disjunctive Datalog. 
This is similar to our definition of declarative
semantics in that models  of $\cal K$ are used to transform the set
 of grounded hybrid rules into a set of ground
Datalog rules, not including DL-atoms. However, as the heads of
the hybrid rules may include DL-atoms, the transformation is more
elaborate than our ${\cal P}/M_0$ transformation. Also the semantics
of ${\cal DL}${+}{\it log} is based on stable models of the transformed ground rules,
while our semantics is based on the well-founded semantics of
${\cal P}/M_0$. 
For stratified normal logic programs
the stable model semantics is equivalent to the well-founded semantics \cite{AptB94}.
Thus for stratified sets of rules
${\cal DL}${+}{\it log}
coincides with our approach
(provided the rules are non disjunctive, and without DL-atoms in their heads,
and the external theory satisfies the requirements of
${\cal DL}${+}{\it log}\/).

The proposed reasoning   algorithm 
(NMSAT-${\cal DL}${+}{\it log}) works
bottom-up 
and is based on grounding. Our operational semantics works top-down
and does not require grounding.  Decidability of ${\cal DL}${+}{\it log} is 
achieved by a weak safeness condition.
The condition is similar to that in our approach.
However here it is not needed for decidability
(but for completeness of the operational semantics;
it is also one of alternative sufficient conditions for soundness).

\commenta{
}

The language of Description Logic
Programs (dl-programs) \cite{eiterLST04a} integrates 
OWL DL with Datalog rules with negation. This is done 
by allowing in the bodies 
so called {\em dl-queries} to a given ontology. The queries may locally
modify the ontology.  Two kinds of declarative semantics are considered for
the integrated language. 
The semantics of choice extends   the stable model semantics
of Datalog
with negation%
\footnote{ More recent versions of this work are 
    based on disjunctive Datalog with negation.
    A further extension is {\em HEX-programs} \cite{EiterIST06}, where
    {\em external atoms} are used to model interface with
    arbitrary external computations.
}
 \cite{eiterLST04a}
 but an extension of the well-founded
semantics is also considered \cite{EiterLST04b}.
In both variants of the declarative semantics 
the {\em truth value} of a rule w.r.t.\ to an interpretation
depends on dl-queries in the rule being  {\em logical consequences} of the
respective ontologies. 
This makes the semantics incompatible
with the standard semantics of the
first order logic.
For example consider
rules $P = \{\, p \leftarrow Q_1;\ p \leftarrow Q_2 \,\}$,
where none of DL-formulae $Q_1,Q_2$ is a logical consequence of the
ontology \T, but in each model of \T at least one of them is true.
Then $p$ is a logical consequence of $P\cup\T$,
but will not follow from $(P,\T)$ 
represented as a dl-program.
In contrast to that, our approach is compatible with FOL, in the sense
explained in Section  
\ref{sec:decl.sem}.
In the last example, $p$ follows from the hybrid program \hp.

\commenta{Was: \ 
}

\section{Conclusions}
\label{sec:concl}

We presented a  framework  for integration of normal logic
programs under the well-founded semantics
with first-order theories. 
Syntactically the integration
is achieved by extending the bodies of  normal clauses with 
(certain) formulae of a given external theory, resulting in the
notions of hybrid rule and hybrid program.
It is assumed that the theories are external 
sources of knowledge and are not modified
during the integration. Therefore it 
is required that the predicates of 
the extended normal program and the predicates of
the  external theory are distinct.

The main contributions of this work are:
\begin{itemize}
\item The declarative semantics of hybrid programs,
combining the (3-valued)  well-founded semantics of normal programs
with the (2-valued) logical semantics of the external 
theory.
It combines non-monotonic negation of the well-founded semantics with 
the classical negation of FOL.
It allows non constant function symbols.
It contrast to most of related approaches it is defined for arbitrary
external FOL theories.

The declarative semantics is undecidable, however it is decidable for
Datalog hybrid programs with decidable external theories.
In the special case of a hybrid program \hp
where \P does not include negation, the declarative
semantics is compatible with the semantics of FOL,
in the sense explained  in Section~\ref{sec:decl.sem}.
The semantics makes possible reasoning by cases (cf.\ Ex.\ \ref{ex:hyrules}).

\comment{
}

\item The  operational semantics that describes how 
to answer (not necessarily ground) conjunctive queries.
This includes handling of non-ground negative queries
by combination  of the ideas of 
 CLP~\cite{CLPhandbook2006} and constructive negation of \cite{drabent95}.
The operational semantics can be seen as an extension
of  SLS-resolution with handling of non-ground negative
queries and constraints of the external theory.
It can be implemented by compilation to Prolog \cite{DrabentHM07},
and employing an LP reasoner for the well-founded semantics and a reasoner
for the external theory.  So the implementation requires rather small
efforts thanks to re-using the existing reasoners.
A prototype implementation of this kind is described in \cite{DrabentHM07}.
It allows non constant function symbols,
the external theories are OWL ontologies.
It uses XSB Prolog \cite{XSB} as an LP engine, and can use various OWL
reasoners.  
It is efficient, in the sense that the number of queries to the OWL
reasoner is small.

\item Soundness and completeness results: the operational semantics
was shown to be sound w.r.t.\ the declarative semantics. It is complete
(and independent from the selection rule)
for safe hybrid programs where the  function symbols are constants.

\end{itemize}

The external theory can be a theory of a constraint domain of CLP.
Thus our framework additionally
 provides a method of adding non-monotonic negation to CLP,
in a way generalizing the well-founded semantics.

It contrast to most of related work,
our approach works for arbitrary external FOL theories.
There are no restrictions on the alphabet of function symbols (it may
be finite or infinite).  
We do not impose any conditions on the equality in the external theories,
like unique name assumption (UNA) or freeness axioms (CET).
There are certain requirements  related to the operational semantics;
however we show how an arbitrary external theory  $\T'$  can be
extended to a theory $\T$ satisfying the requirements.

This paper does not study implementation techniques.
They are subject of future work, begun in \cite{DrabentHM07}.

\paragraph*{Acknowledgement.}
This research has been partially funded by the European Commission and by the
Swiss Federal Office for Education and Science within the 6th
Framework Programme project REWERSE number 506779
(cf. \url{http://rewerse.net}).

\appendix
\section*{Appendix}

Here we present proofs of  Lemma \ref{lemma.safe}
and of two main technical lemmas from 
Sections \ref{sec:soundness} and~\ref{sec:completeness}.
By $var(F)$ we denote the set of free variables of a formula $F$.

\paragraph{Proof of Lemma  \ref{lemma.safe}.}

We first prove part 2 by induction.  Then part 1 follows immediately.
Assume that a node $G_i$ is safe apart of $V$.  We show that 
each its child is safe apart of $V$.

\newcommand*{\olnn}{\ensuremath{\overline{N'}}}
\newcommand*{\oln}{\ensuremath{\overline{N}}}

Let $G_i=C,\oll,p(\ol t),\olll$ and its child
$G_{i+1} = \ol t\eq \ol u, C, C', \oll,\olk,\olll$
be derived from $G_i$ by a rule $R=p(\ol u)\gets C',\olk$.
Let $\ol N$ be the negative literals of $\oll,\olll$, and
 $\ol{N'}$ be the negative literals of $\olk$.
Each free variable of $\ol u, C',\olnn,$ is bound in $C'$ to a ground term or to a
variable occurring in a positive literal of  \olk\ 
(as $R$ is safe).
Each variable from $var(C,\oln)\setminus V$
is bound in $C$ to a ground term or to a
variable occurring in a positive literal of $\oll,\olll$ or in $\ol t$
(as $G_i$ is safe apart of $V$).
Hence each variable from
$var(C,\oln,\ol t\eq \ol u, C',\olnn)\setminus V$
is bound in  $C,\ol t\eq \ol u, C'$  to a ground term or to a
variable occurring in a positive literal of $\olk,\oll,\olll$.

Let $G_i=C,\oll,\neg A,\olll$ with a negative literal $\neg A$
selected, with a child $G_{i+1} =  C, C', \oll,\olll$.
We have $var(C')\subseteq var(C,A)$, thus
$var(C,C')\setminus V\subseteq var(C,A) \setminus V$. 
Thus each variable from
$var(C,C')\setminus V$ is bound in  $C, C'$  to a ground term or to a
variable occurring in a positive literal of $\oll,\olll$.
As the negative literals of $G_{i+1}$ are those of $G_i$, 
we obtain that  $G_{i+1}$ is safe apart of $V$.
\hfill $\Box$
\medskip

The proofs below refer to
ground programs of the form $P/M_0$ (cf.\ Def.\,\ref{def.declarative}),
and the operator defining the well-founded
semantics of normal logic programs (cf.\ Section~\ref{sec:wfs}):
\[
\psipm(I) = (
 {\cal M}_\PMtI  \cap {\cal H}
 	    )  \cup  \neg ( {\cal H} \setminus
 	{\cal M}_\PMtuI
 			   )
\]
Remember that from monotonicity of \psipm it follows that
$k'\leq k$ implies
$\psipmpower{k'}\subseteq\psipmpower{k}$.

\comment{
}

\paragraph{Proof of Soundness Lemma  \ref{lemma.soundness}.}
\label{lemma.soundness.proof}

By transfinite induction on the rank of the t-tree and tu-tree.
Assume that the lemma
holds for t-trees and tu-trees of rank $<k$.
We can assume that $\theta$ binds only the free variables of $G$.

\comment{}

1. Assume that $C$ is an answer of a t-tree for $G$ of rank $k$,
and that   $M_0\models C\theta$.
Consider the branch  $\seq[m]G$ of the tree from the root $G$ to the leaf
 $C'$ (i.e.\ $G_1=G$, $G_m=C'$)
such that $C=\restrict{(\ldots\lor C'\lor\ldots)}G$
and $M_0\models(\restrict{C'}G)\theta$.
We show that some ground instance of $C'\theta$ is true in $M_0$.
This is obvious when $M_0$ is a Herbrand interpretation or when the set of
constraints has the witness property.
If $P$ is safe then
$C'$ is safe apart from the set $V_0=var(G)$,  by Lemma \ref{lemma.safe}.
So each variable $x\in var(C')\setminus V_0$ is bound in $C'$ to a ground
term $t_x$. 
Consider a substitution
 $\varphi = \{\, x/t_x \mid x\in var(C')\setminus V_0\,\}$.
Formula $\restrict{C'}G$ is equivalent to $\restrict{(C'\varphi)}G$
and to ${C'\varphi}$.
Hence $M_0\models C'\varphi\theta $.
The latter formula is a ground instance of $C'\theta $, as
  $\varphi\theta = \theta\varphi$.

\commenta{
safe etc referred to only here
}

Thus
substitution $\theta$ can be extended to a substitution $\theta'$ such that
$M_0\models C'\theta'$, and $G\theta'=G\theta$.
Notice that
$\theta'$ is applicable to all the goals in the branch
(as for any such goal its constraint is of the form $\seq[l]C$, where
 $C'=\seq C,\ l\leq n$).

The sequence $G_1\theta',\ldots,G_m\theta'$ is a successful branch of
an instance of the considered t-tree for $G$.
If a positive literal is selected in $G_i$ then $G_{i+1}\theta'$ is
derived from  $G_i\theta$  by a rule from $(P/M_0)$.
Whenever a negative literal $\neg A$ is selected in $G_i$
and $C_i$ is the constraint of $G_i$
then the constraint of  $G_{i+1}$ is $C_i,C''$,
where $C''$ is a negative answer of a tu-tree of rank $j<k$ for $C_i,A$.
As $\theta'$ is applicable to $G_{i+1}$, we have
$M_0\models (C_i,C'')\theta'$.
By the inductive assumption, 
$\psipmpower{j+1}\models_3 \neg A\theta'$.
The same holds for any $k'\geq j$.

Thus the sequence $G_1\theta',\ldots,G_m\theta'$ is a successful
SLD-derivation for $G\theta'$ and the program $P_t=\PMt{\psipmpower{k'+1}}$
for some $k'<k$.
(If $k=0$ then  $P_t=P/M_0$.)
  Hence  ${\cal M}_{P_t}\models G\theta$,
by soundness of SLD-resolution \cite{Apt-Prolog}.
By the definition of $\psipm$,
for each positive literal $A'$ of $G\theta$ we have 
$\psipmpower{k'+2}\models_3 A'$.
For each negative literal $\neg A'$ of $G\theta$ we have 
$\neg A'\in P_t$, i.e.\  $\psipmpower{k'+1}\models_3\neg A'$.
 By monotonicity of \psipm for each literal $L$ of $G\theta$ we have 
$\psipmpower{k+1}\models_3 L$.  Hence 
$\psipmpower{k+1}\models_3 G\theta$.

\smallskip
2. Consider a tu-tree $\Tr$ of rank $k$ for $G$.
Let the negative answer $C$ be obtained from $\Tr$ and its cross-section $F$.
Let $\it CF$ be the constraints of the nodes of $F$.
So $C$ implies $\neg(\restrict{C_i}G)$
for each $C_i\in CF$.
Assume that $M_0\models C\theta$.
Then  $M_0\notmodels C_i\theta'$, for any extension $\theta'$ of $\theta$
grounding the variables of $C_i$ and for each $C_i\in CF$.  Thus,
in a ground instance $\Tr\theta$ of the tree,
no node corresponds to a node of $F$
(nor to a descendant of a node of $F$).
Hence no leaf of  $\Tr\theta$ is successful.

Out of $\Tr\theta$ we construct an SLD-tree with root $G\theta$ for a certain
ground program.  The tree has no success leaves, hence $G\theta$ is false
in the least Herbrand model of the program (by completeness of SLD-resolution).

In $\Tr\theta$,
consider a leaf $G'\theta'$ corresponding to a node
$G' = (C',\ol L,\neg A,\ol{L'})$ of $\Tr$, with a negative literal
$\neg A$ selected. 
The node $G'$ does not have a child 
$C',\ol L,\ol{L'}$
(such a child implies that $G'\theta'$ is not a leaf).
Hence case \ref{sls.negation.tu.child} or \ref{sls.negation.tu.leaf}
has been applied to the node.
Thus there exists an answer  $C''$ of a t-tree
for $C',A$ of rank $k'<k$.
Moreover $(\neg C'',C')\theta'$ is false in $M_0$
(otherwise $\neg C'',C'$ is satisfiable, $G'$ has a child
$\neg C'',C',\ol L,\ol{L'}$,
and $G'\theta'$ has a child $\ol L\theta',\ol{L'}\theta'$).
Hence $\neg C''\theta'$ is false in $M_0$ (as $C'\theta'$ is true in $M_0$).
From $M_0\models C''\theta'$ and from the inductive assumption we obtain
$\psipmpower{k'+1}\models_3 A\theta'$.
By monotonicity of \psipm
the same holds for any $k''\geq k'$.
Thus $\neg A\theta'\not\in \PMtu{\psipmpower{k'+1}}$,
by the definition of $/_{tu}$.
As $k'+1\leq k$, we have 
$\neg A\theta'\not\in \PMtu{\psipmpower{k}}$.
Hence the node  $G'\theta'$ (with  $\neg A\theta'$ selected)
is a leaf of an SLD-tree for program $\PMtu{\psipmpower{k}}$.
(Notice that this reasoning is also valid for a non finite $k$.
In particular, $k'+1< k$ if $k$ is a limit ordinal.)

If a non-leaf node $G'\theta'$ in $\Tr\theta$ corresponds to 
$G' = (C',\ol L,\neg A_{G'},\ol{L'})$ of $\Tr$,
with $\neg A_{G'}$ selected, then the child of  $G'\theta'$ is
$(\ol L,\ol{L'})\theta'$.
For each such node $G'\theta'$ of $\Tr\theta$, let us add $\neg A_{G'}\theta'$
to the child and all its descendants.
The obtained tree is an SLD-tree for $G\theta$ and
a program  $P_{tu}\cup P_{\rm void}$,
where  $P_{tu} =\PMtu{\psipmpower{k}}$,
and each clause of $P_{\rm void}$ is of the form $\neg A\gets\neg A$.
Adding $P_{\rm void}$ to a definite clause logic program does not change its
least Herbrand model.
The obtained tree does not have a success node.
By completeness of SLD-resolution
\cite{Apt-Prolog}
${\cal M}_{P_{tu}}\notmodels G\theta$, and for some literal $L$ of $G\theta$
we have $L\not\in{\cal M}_{P_{tu}}$.

If $L$ is of the form $\neg A$ then 
$A\in {\psipmpower{k}}$ (by the definition of $/_{tu}$),
and $L$ is \fff in ${\psipmpower{k}}$.
Otherwise 
$L\in {\cal H}\setminus{\cal M}_{P_{tu}}$ and $\neg L\in {\psipmpower{k+1}}$,
hence $L$ is \fff in ${\psipmpower{k+1}}$.
In both cases $L$ is \fff in $\psipmpower {k+1}$,
thus $\psipmpower {k+1}\models_3\neg G\theta$. \hfill $\Box$

\comment{}

\paragraph{Proof of Completeness Lemma \ref{lemma.completeness}.}
If  $\psipmpower k\models_3 G\theta$ or
 $\psipmpower k\linebreak[3]\models_3\neg G\theta$
then we say that $k-1$ is a level of $G\theta$ when $G$ does not contain a
negative literal 
and $k$ is a level of $G\theta$ when $G$ contains a negative literal.
The proof is by induction on a level of $G\theta$;
the rank of the constructed t- or tu-tree for $G$ is the level of $G\theta$.

Let  $G\theta$ be of level $k'\in\{k-1,\,k\}$.
Let us denote $I= \psipmpower{k-1}$
and  $J = \psipmpower{k'}$.  So $J=\psipm(I)$ if $G$ contains a
negative literal, and $J=I$ otherwise.  In both cases  $I\subseteq J$.

\medskip
1.  
Assume $\psipmpower{k}\models_3 G\theta$.
Each positive literal $A$ in $G\theta$ is a member of $ {\cal M}_\PMtI$
(by the definition of $\Psi$) and hence of ${\cal M}_\PMt J$.
For each negative literal $\neg A$ in $G\theta$,
we have $J \models_3 \neg A$ (as then $\psipmpower{k} = J$).
Thus $\neg A\in \PMt J$, by the definition of $/\!_t$.
So, by completeness of SLD-resolution, for any selection rule
there exists a successful
SLD-derivation $\cal D$ for the goal $G\theta$ and program  $ \PMt J$.

We now show how negative literals in the derivation are related to lower
rank tu-trees.
Consider a negative literal  $\neg A$ occurring in $\cal D$.
We have $\neg A\in \PMt J$
(i.e.\ $\neg A\in J$), so  $A$ is of level $k'-1$.
By the inductive assumption, 
for any goal $B$ such that $A=B\theta'$
is a ground instance of $B$,
if $C'$ is the negation of a finite pseudo-answer 
of a maximal tu-tree for $B$ via $\R$ of rank $k'-1$
then
$M_0\models C'\theta'$.

Consider a maximal t-tree $\Tr$ of rank $k'$ for $G$ and $\hp$ via \R.
Let $\Tr\theta$ be a ground instance of $\Tr$.
It can be seen as an SLD-tree via a selection rule $\R'$,
for a certain ground definite program containing the rules of $P/M_0$
and some facts of the form $\neg A$.  For $\R'$ there exists a derivation 
$\cal D$ as above.
We show that $\cal D$ is a branch of $\Tr\theta$.
We prove by induction on $i$ that
the $i$-th goal of $\cal D$ is a node of $\Tr\theta$.
Let a goal $G'\theta'$ of $\cal D$ be a node of $\Tr\theta$,
corresponding to a node $G'$ of $\Tr$.  
If a positive literal is selected in $G'\theta'$ and in $G'$
then any goal derived from  $G'\theta'$ by a rule from $P/M_0$
is a ground instance $G''\theta''$ of a child $G''$ of $G'$ in $\Tr$,
where $\theta''$ is an extension of $\theta'$.
If a negative literal $\neg A$ is selected in $G'\theta'$
then a $\neg B$ is selected in $G'$ and $\neg A=\neg B\theta'$.
The tree $\Tr$ refers to a maximal tu-tree for $B$,
and to the negation $C'$ of its finite pseudo-answer.
As explained in the previous paragraph,
  $M_0\models C'\theta'$ follows from the inductive
assumption. 
The child $G''$ of $G'$ in $\Tr$ is $G'$ with $\neg B_i$ removed and
$C'$ added.
So substitution $\theta'$ is applicable to $G''$,
and
$G''\theta'$ is $G'\theta'$ with $\neg A_i$ removed.  Thus the successor
of $G'\theta'$ in $\cal D$ is the child $G''\theta'$ of $G'\theta'$ in $\Tr\theta$.

The last goal of $\cal D$ is empty, it is a ground instance of a leaf 
$C_\mathrm{s}$ of $\Tr$.  Thus $M_0\models \restrict{(C_\mathrm{s}\theta)}G$.
Let $C$ be a finite answer of the t-tree.  $M_0\models C\theta$.

\medskip
2. Assume that $\psipmpower{k}\models_3\neg G\theta$.
Some literal $L$ of $G\theta$ is
false in $\psipmpower{k}=\psipm(I)$.
If $L=A$ is an atom then $\neg A\in\psipm(I)\subseteq\psipm(J)$.
Hence $A\not\in{\cal M}_\PMtu{J}$ 
(by the definition of \psipm).
If $L=\neg A$ is a negative literal then $A\in\psipm(I)=J$.
Hence $\neg A\not\in \PMtu J$ (by the definition of $/\!_{tu}$).
In both cases, $L\not\in{\cal M}_\PMtu{J}$.
Thus, by soundness of SLD-resolution,  
there does not exist a successful SLD-derivation
for $G\theta$ and the program $\PMtu J$
(as otherwise $L\in{\cal M}_\PMtu{J}$, contradiction).

Consider a maximal tu-tree $\Tr$ of rank $k'$
for $G$ via $\R$, with a pseudo-answer $D$.
We have to show that $M_0\models\neg D\theta$.
Assume the contrary; then $\Tr$ has a success leaf $C_\mathrm{s}$ such that 
$M_0\models(\restrict{C_\mathrm{s}}G)\theta$.
Consider a branch of $\Tr$ from $G$ to $C_\mathrm{s}$.
Let  $\seq[l]G$ be the goals of the branch ($G=G_1$, $G_l=C_\mathrm{s}$).
As $P$ is safe, by Lemma  \ref{lemma.safe}
there exists a substitution $\theta'$ which is an extension of $\theta$
and is applicable to the nodes of the branch.
(See part 1 of the proof of Lemma \ref{lemma.soundness} for details.)
We show that $G_1\theta',\ldots,G_l\theta'$ is a successful
SLD-derivation for $G\theta$ and
\PMtu J, which is a contradiction.

\commenta{
Safeness used only here.
}

If a positive literal is selected in $G_i$ and a rule $r\in P$
is applied to $G_i$ to obtain $G_{i+1}$
then $G_{i+1}\theta'$ is obtained from  $G_{i}\theta'$ by applying 
a normal ground instance $r\sigma\in \PMtu J$ of $r$.
Assume a negative literal $\neg B$ is selected in $G_i$.
The constraint of $G_{i+1}$ is equivalent to $C_i,\neg C''$ where 
$C_i$ is the constraint of $G_i$ and $C''$ is a finite answer of a
maximal t-tree of rank  $k'-1$
for $B$.  
We have  $M_0\models\neg C''\theta'$,
as $\theta'$ is applicable to $G_{i+1}$.
Suppose $B\theta'\in J$.
This leads to contradiction:
such $B\theta'$ is of level $k'-1$ and, by the inductive assumption,
the answer $C''$ satisfies $M_0\models C''\theta'$.
Thus $B\theta'\not\in J$ and  $\neg B\theta'\in \PMtu J$.
Thus  $G_{i+1}\theta'$ is obtained from  $G_{i}\theta'$ by applying 
a rule from \PMtu J.

Hence $G_1\theta',\ldots,G_l\theta'$ is a successful
SLD-derivation for $G\theta$ and \PMtu J; the assumption that 
$M_0\models(\restrict{C_\mathrm{s}}G)\theta$ is false.
So $M_0\models\neg(\restrict{C_\mathrm{s}}G)\theta$
for each success leaf $C_\mathrm{s}$ of $\Tr$.
Let $D$ be a finite pseudo-answer of $\Tr$.
Then $M_0\models \neg D\theta$ for a negative answer $\neg D$ of the tree.
\mbox{}\hfill
$\Box$

\bibliographystyle{alpha} 
\bibliography{biblio.DM}

\end{document}